\documentclass[trackchanges]{aastex7}

\shorttitle{Simulations of Secondary Heliospheric Current Sheet Formation}
\shortauthors{Casillas et al.}

\begin{document}


\title{Bifurcations in the Heliosphere: Global-scale Coronal Pseudostreamer Evolution\\ and the Creation of Secondary Heliospheric Current Sheets}

\author[0000-0001-6248-183X]{Lizet Casillas}
\affiliation{Department of Earth, Planetary, and Space Sciences, University of California--Los Angeles, Los Angeles, CA 90056, USA}
\email[show]{lizetcasillas@ucla.edu}  

\author[0000-0001-6886-855X]{Benjamin~J.~Lynch} 
\affiliation{Department of Earth, Planetary, and Space Sciences, University of California--Los Angeles, Los Angeles, CA 90056, USA}
\email{bjlynch@ucla.edu}  

\author[0000-0002-2916-3837]{Victor R\'eville}
\affiliation{Univ Toulouse, CNES, CNRS, IRAP, Toulouse, France}
\email{victor.reville@utoulouse.fr}  

\author[0000-0002-4440-7166]{Olga~Panasenco}
\affil{Advanced Heliophysics, Monrovia, CA 91016, USA}
\email{panasenco.olga@gmail.com}

\author[0000-0002-2381-3106]{Marco Velli}
\affiliation{Department of Earth, Planetary, and Space Sciences, University of California--Los Angeles, Los Angeles, CA 90056, USA}
\email{mvelli@ucla.edu}

\begin{abstract}
%

The heliospheric current sheet (HCS) is the boundary between open magnetic fields in the solar wind of opposite polarity that connect back to their respective open flux regions in the corona. The HCS is often described as a disc-like sheet warped by the combined effects of the inclination of the magnetic equator on the Sun, solar rotation, and solar wind expansion. 
The HCS is an extension of the helmet streamer belt---the closed-flux region resulting from the global dipole component of the observed photospheric magnetic field. In the absence of a strong dipole, a dominant quadrupole can result in two separate HCSs.
Here we report on a set of idealized, axisymmetric solar wind simulations designed to study the transition of the single-HCS corona and inner heliosphere into a multiple-HCS configuration, run with the WindPredict-AW PLUTO MHD code. 
We systematically vary the relative strengths of the dipole/quadrupole contributions to the global field and show this results in the formation of a large-scale, closed-flux pseudostreamer that interacts with the ambient solar wind. 
We find that when the dipole and quadrupole magnitudes are approximately equal, our model corona can oscillate between a one- and two-HCS configuration on the timescale of days-to-weeks. 
We characterize the plasma dynamics of the transition of a global-scale pseudostreamer into a localized, secondary helmet-streamer belt with its own HCS surrounding a newly-opened, opposite-polarity coronal hole. 
We discuss the implications of our results for solar cycle evolution and the dynamic coupling of the highly structured corona and inner heliosphere.

\end{abstract}


\section{Introduction} 
\label{sec:intro}

As the solar corona expands to become the solar wind it drags the solar magnetic field into interplanetary space. When closed magnetic field lines, with footpoints rooted in positive and negative polarity regions of photospheric magnetic flux, expand and/or are swept with the accelerating wind beyond the Alfv\'{e}n critical point to super-Alf\'{e}nic velocities ($M_A \ge 1$) they can be considered ``open'' and part of the interplanetary magnetic field \citep{McComasetal07, Badman2025}. The heliospheric current sheet (HCS) is the boundary between the oppositely directed magnetic field lines. The HCS is almost always embedded in a thicker heliospheric plasma sheet \citep[HPS;][]{Winterhalter94}---a single continuous plasma structure with enhanced electric current density layers and large-scale latitudinal excursions above and below the ecliptic plane, resembling a so-called ``ballerina skirt'' heliospheric geometry and a local thickness (width) of approximately 10,000 km at 1~AU \citep{smith_heliospheric_2001,Mursula12}.  
The coronal origin of the HCS/HPS and the adjacent open--closed boundary regions of large-scale closed coronal flux systems serve as a primary source of slow solar wind \citep{WangYM2000} and the HCS/HPS can significantly influence the overall structure of the heliosphere \citep{Riley2012}.  The HCS/HPS plays a critical role in space weather phenomena; in addition to its impact on various plasma parameters, such as solar wind speed, density, temperature \citep[e.g.,][]{Abbo2016}, it can also influence the trajectory of coronal mass ejections (CMEs) via deflection, distortion and other interaction processes \citep[e.g.,][]{Manchester2017,Luhmann2020}. 
Recent observational and numerical simulation studies have produced various interpretations of HCS structure and dynamics, each with a common theme: the smooth, singular, time-stationary description of the HCS/HPS ``disc'' cannot account for the considerable slow solar wind variability observed in situ. 
The HCS/HPS is a broad region composed of a complex succession of density blobs and small-scale magnetic flux ropes \citep{kepko2016,higginson_lynch_2018,sanches-diaz2019}. 
Parker Solar Probe \citep[PSP;][]{Fox2016} observations with the Wide-field Imager for Solar Probe \citep[WISPR;][]{Vourlidas2016} instrument have imaged coronal streamer structure, dynamics, and evolution with unprecedented detail, including the formation of coronal inflows and in/out pairs \citep{Vourlidas2025,Liewer2024} and the internal structure of CMEs and streamer-blob flux ropes \citep{Cappello2024,Ascione2024}. These small-scale streamer-blob flux rope-like transients appear to be superimposed or embedded with the larger HCS/HPS structure, which itself can include fairly significant mass density gradients associated with the fine-scale, collimated rays emanating from the helmet streamer cusps into the HCS/HPS \citep{Liewer2023}. 
Combining multi-spacecraft data with 3D MHD simulations, \citet{reville_flux_2022} modeled the origin of flux-ropes in slow solar wind during the June 2020 conjunction of Parker Solar Probe  and Solar Orbiter \citep[SolO;][]{Mueller2020}. These simulation results confirm that sequential magnetic reconnection at the tip of helmet streamers can produce periodic flux rope ejection into the HCS.


Pseudostreamers are closed coronal magnetic structures sitting above parasitic-polarity flux regions surrounded by open fields lines of the opposite (dominant) polarity \citep{wang_ps_2007, titov2011, titov2012, Panasenco2013, Rachmeler2014}. 
A separatrix surface encloses the domes separating them from open magnetic fields. 
These configurations are often sites of interchange reconnection. The coronal null point and the susceptibility of electric current density accumulation on the pseudostreamer's separatrix boundary in response to energization and system stresses creates favorable conditions for the onset of magnetic reconnection \citep[][and references therein]{Antiochos1990, Demoulin1996, Edmondson2017, Wyper2022}. The dynamic interaction between closed and open fields at these topological domain boundaries has been shown to produce a variety of structured, reconnection-generated transients, from density enhancements and intermittent, jet-like outflows to a spectra of low-frequency Alfven waves \citep{WangYM2008, WangYM2012, WangYM2019, Lynch2013, Lynch2014, Lynch2023, Masson2014, Gannouni2023, Pellegrin-Frachon2023, Wyper2022, Romano2025}.



The solar dynamo, operating deep in the interior, generates the solar magnetic field and thus, ultimately drives all solar activity \citep{Charbonneau2020}. The approximately 11-year cycle of sunspot (strong field active region) emergence and spatiotemporal evolution and the $\sim$22-year magnetic cycle where the strength (and polarity) of the polar magnetic fields exhibit a fairly robust sinusoidal-like time-dependence \citep{Hathaway2015}. Solar minimum refers to the interval where the polar fields are at their maximum and the sunspot number is at its lowest, while solar maximum occurs during the interval of greatest number sunspots and polar sign reversal (polar field strengths $\sim$ zero). 
The solar wind is also directly correlated to the Sun’s solar cycle, with maxima experiencing more solar wind activity than the solar minima. During solar minimum, a single ``band'' of closed-flux (the helmet streamer belt) encircles the Sun. The equatorial helmet streamer belt dominates the coronal magnetic structure during solar minimum, while at solar maximum, the dipole-like coronal structure gives way to quadrupole dominance \citep{Wang2014SSR}. Given that the structure of the current sheet is an extension of the solar magnetic field, it is expected that the coronal structure will undergo changes as the Sun's magnetic field transitions to higher-order multipoles \citep{WangYM2003}. The global (average) structure of the HCS is expected to exhibit a disc-like configuration during solar minimum arising from an almost purely dipolar field, while at solar maximum, the magnetic field's quadrupole dominance generates two, conical HCSs \citep{sokoloff_symmetries_2020}. In fact, \citet{wang_evidence_2014} presented observations and PFSS magnetic field extrapolations of such a quadrupole-dominated period to show a bifurcated heliosphere with two global-scale conical HCS structures during mid-2012 as the polar fields were undergoing their reversal during Solar Cycle 24.

To gain better understanding of the coronal origins of HCSs and how their formation and evolution shape the heliosphere, we perform a set of axisymmetric MHD solar wind simulations to investigate how the global coronal flux systems---arising from photospheric magnetic flux distributions---respond to the presence of a quasi-steady state solar wind outflow. 
%
%
%
%
%
The structure of the paper is as follows. In section 2, we present a description of the numerical MHD code used to perform our set of axisymmetric solar wind relaxation runs, as well as the procedure for constructing the initial magnetic field configuration and selection of the solar wind outflow parameters. In section 3, we discuss our simulations results in detail, starting with the field and plasma evolution of the opening-up and closing-down of magnetic flux during the pseudostreamer-to-helmet streamer and helmet streamer-to-pseudostreamer transition process, the influence of the solar wind model on our results, and quantifying the relative magnitudes of the dipole and quadrupole components in the ``unstable'' pseudostreamer transition regime. In section 4, we discuss the implications of our results for the gradual reconfiguration and evolution of global-scale corona, as well as outline the next steps for further modeling and observations of pseudostreamer-to-helmet streamer transitions that form long-lived secondary HCS/HPS structures surrounding new ``coronal funnel'' solar wind streams.

\section{Methods} \label{sec:methods}

\subsection{AW-Predict Extension of the PLUTO MHD Code} \label{subsec:plutocode}

To investigate the heliospheric consequences of the structure of the global corona, the distribution of its open- and closed-flux regions, and the resulting locations of HCSs/HPSs, we use the WindPredict-AW version \citep{reville_2020b} of the open-source, compressible MHD code, PLUTO \citep{mignone_2007}. 
The WindPredict-AW numerical simulations are performed in axisymmetric (2.5D) spherical coordinates with the usual $z$-axis of azimuthal symmetry that eliminates the ${\phi}$-coordinate dependence from the MHD system. Thus, all vector quantities retain their $\mathbf{\hat{\phi}}$ components but only vary with respect to the $(r,\theta)$ coordinates.
The MHD equations are solved using an improved
Harten–Lax–van Leer (HLLD) Riemann solver \citep{Miyoshi2005} along with slope-limited parabolic reconstruction, while the solenoidal condition, $\nabla \cdot \mathbf{B} = 0$, is enforced through the hyperbolic divergence cleaning described in \citet{Dedner2002}.

WindPredict-AW/PLUTO uses this reconstruct–solve–average approach to solve the following equations of MHD and Alfv\'{e}nic wave energy transport:
\begin{equation} \frac{\partial}{\partial t}\rho + \nabla \cdot \left( \rho \mathbf{v} \right) = 0 \;, \end{equation}
\begin{equation} \frac{\partial}{\partial t}  \rho v + \nabla \cdot  \left( \rho v \mathbf{v} - \mathbf{B}\mathbf{B} + \mathbb{I}{p} \right) = - \rho \nabla \Phi_g  \; , 
\end{equation}
\begin{equation} \frac{\partial}{\partial t} \left( E + \mathcal{E}_{\rm AW} + \rho \Phi_g \right) + \nabla \cdot \left[ \left( E + p + \rho\Phi_g \right) \mathbf{v} - \mathbf{B}\left( \mathbf{v} \cdot \mathbf{B} \right) + \mathbf{v}^{+}_{\rm AW} \mathcal{E}^{+}_{\rm AW} + \mathbf{v}^{-}_{\rm AW} \mathcal{E}^{-}_{\rm AW} \right] = Q \; ,
\label{mhd:energy}
\end{equation} 
\begin{equation} \frac{\partial}{\partial t}\mathbf{B} + \nabla \cdot (\mathbf{v}\mathbf{B}-\mathbf{B}\mathbf{v}) = 0 \; .
\end{equation} 
The variables retain their usual meaning, e.g., $\rho$ is the mass density, $\mathbf{v}$ is the velocity field, $\mathbf{B}$ is the magnetic field, $\mathbb{I}$ is the identity matrix, $p$ is the total pressure, and $\Phi_g$ is the gravitational potential $\Phi_g = G M_\odot/r$ which yields solar gravitational acceleration $\mathbf{g}_\odot = -\nabla \Phi_{g}$.
The total pressure given as the sum of the thermal pressure, Alfv\'{e}n wave pressure, and magnetic pressure, 
\begin{equation}
    p=p_{\rm th} + \frac{\mathcal{E}_{\rm AW}}{2} + \frac{B^2}{8\pi} \; . 
\end{equation}
In Equation~\ref{mhd:energy}, the energy density $E$ represents the internal, kinetic and magnetic contributions, 
\begin{equation}
E = \rho e + \frac{\rho v^2}{2} + \frac{B^2}{8\pi} \; .
\end{equation}
The internal energy density is obtained via the ideal equation of state $\rho e = p_{\rm th}/\left( \gamma-1\right)$ with the ratio of specific heats $\gamma=5/3$.
The total Alfv\'{e}nic wave-turbulence energy density is given by $\mathcal{E}_{\rm AW}$ which is itself the sum of waves propagating along ($-$) and against ($+$) the magnetic field, $\mathcal{E}_{\rm AW} = \mathcal{E}_{\rm AW}^{+} + \mathcal{E}_{\rm AW}^{-}$. Note that for open field lines we consider explicitly only waves propagating outwards from the Sun, and the two populations are never present together, rather when the magnetic field orientation is outward, we have 
$\mathcal{E}_{\rm AW}^{-}$ and if the field is inward, then we have $\mathcal{E}_{\rm AW}^{+}$.
This approximation has been relaxed in \citet{Reville2018}.
The Alfv\'{e}n wave group velocities are defined as the usual $\mathbf{v}_{\rm AW}^{\pm} = \mathbf{v} \pm \mathbf{v}_A$. 
The energy equation source term $Q = Q_h - Q_c - Q_r$ represents the sources and sinks due to coronal heating $Q_h$, thermal conduction $Q_c$, and radiative losses $Q_r$. The coronal heating is taken as an exponential
\begin{equation}
Q_h = \frac{F_h}{H}\left( \frac{R_\odot}{r} \right)^2 \exp{\left[ - \frac{\left(r-R_\odot\right)}{H} \right]}
\end{equation}
with a characteristic scale-height of $H$ and a photospheric energy flux of $F_h$. 
The implementation of the thermal conduction $Q_c$ follows the typical practice of a smooth transition from the Spitzer-H\"{a}rm collisional heat flux prescription in the low corona $r < r_{\rm coll}$ to the collisionless regime further out $r > r_{\rm coll}$ given by
\begin{equation}
Q_c = - \nabla \cdot \left[ \, \alpha(r) \, \mathbf{q}_S + \left( 1 - \alpha(r) \right) \mathbf{q}_P \, \right] \;\; ,
\end{equation}
where $\mathbf{q}_S = -\kappa_0 T^{5/2} \nabla T$, 
$\mathbf{q}_P = (3/2)\, p_{\rm th} \mathbf{v}$, and
$\alpha(r) = \left( 1 + [( r - R_\odot)/( r_{\rm coll} - R_\odot)]^4 \right)^{-1}$
\citep[see, e.g.,][]{Hollweg1986}.
The radiative losses are given by the form
$Q_r = n^2 \Lambda (T)$ 
where $n = n_p = n_e = (\rho/m_p)$ is the proton or electron number density assuming charge neutrality and radiative loss function $\Lambda(T)$ given by \citet{Athay1986}. 

The Alfv\'{e}nic wave-turbulence energy densities $\mathcal{E}^{\pm}_{\rm AW}$ and their local dissipation $Q^{\pm}_{\rm AW}$ can be written concisely with the Els\"{a}sser variable $\mathbf{z}^{\pm}$ notation as
\begin{equation}
\mathbf{z}^{\pm} = \delta \mathbf{v} \mp \frac{\delta \mathbf{b}}{\sqrt{4\pi \rho}} \;, \;\;\;\;
\mathcal{E}^{\pm}_{\rm AW} = \frac{\rho}{4} |\mathbf{z}^{\pm}|^2 \;, \;\;\;\; Q^{\pm}_{\rm AW} = \frac{\mathcal{E}^{\pm}_{\rm AW}}{2\lambda} \left( \, |\mathbf{z}^{\mp}| + \mathcal{R} \, |\mathbf{z}^{\pm}| \, \right) \;\; ,
\end{equation}
%
where $\delta\mathbf{b} = \mathbf{B}-\langle\mathbf{B}\rangle$ is the fluctuation component defined with respect to the average background field $\langle \mathbf{B} \rangle$, the local wave energy dissipation scale-length $\lambda$ is taken a function of the magnetic field strength $\lambda = \lambda_\odot / \sqrt{B}$, and a small reflection coefficient $\mathcal{R}=0.1$ is added to mimic dissipation in the open-field regions \citep{Verdini2007,vanderHolst2010}.
The transport equations for the Alfv\'{e}nic wave-turbulence energy densities in the WKB approximation \citep{Alazraki1971,Belcher1971} are thus,
\begin{equation}
\frac{\partial}{\partial t} \mathcal{E}^{\pm}_{\rm AW} + \nabla \cdot \left( \mathbf{v}^{\pm}_{\rm AW} \mathcal{E}^{\pm}_{\rm AW} \right) = - \frac{\mathcal{E}^{\pm}_{\rm AW}}{2} \nabla \cdot \mathbf{v} - Q^{\pm}_{\rm AW} \; ,
\end{equation}
which are advanced in time alongside the MHD equations of mass, momentum, and energy density transport and the induction equation throughout the computational domain \citep{reville_2020b}. 

%

In the next section we briefly summarize the numerical modeling setup used in our set WindPredict-AW runs, including the computational grid and the parameter values used to describe the simulations' coronal magnetic field, solar atmosphere, and coronal heating/solar wind models.
 
\subsection{Numerical Simulation Setup and Parameter Values} \label{subsec:simparameters}

Each of our WindPredict-AW simulations are run with an identical numerical grid. The $r$--$\theta$ computational domain is $r \in [1 R_{\odot}, 20 R_{\odot} ]$, $\theta \in [0, \pi]$ with $128 \times 128$ points used in the computational grid. 
We employ the \citet{reville_2020b} strategy for the  grid decomposition in the $r$-direction where the grid is a combination of a $10 \times 128$ uniform boundary layer with $(\Delta r, \Delta \theta) = (0.0004 R_\odot, 0.0078\pi)$ and a $118 \times 128$ geometrically-stretched grid with, for example  $\Delta r/r=(0.058, 0.063, 0.066)$ at $ r = (5.54,10.6, 15.5)$
respectively, and $\Delta \theta $ always remains constant at $0.0078\pi$

%


The 3D magnetic field in the code can be specified either directly as the $g_{lm}, h_{lm}$ coefficients of the spherical harmonic expansion of a potential field.  In the current study, we construct our initial coronal field directly from the spherical harmonic coefficients corresponding to the axisymmetric dipole ($g_{10}$; $\ell=1$, $m=0$) and quadrupole ($g_{20}$; $\ell=2$, $m=0$) terms as the azimuthal symmetry in our 2.5D system prohibits the $m \ne 0$ values. 
Our initial ($t=0$) 3D magnetic field configurations are therefore given by   
\begin{eqnarray}
B_{r}(r,\theta) &=& \sum_{\ell=1}^{2}  \, g_{\ell0} \, \left( \frac{R_\odot}{r} \right)^{\ell+2} 
\, P_{\ell}^{0}( \cos{\theta} ) \,\,\, ,\\
B_{\theta}(r,\theta) &=& \sum_{\ell=1}^{2} \, g_{\ell0} \, \left( \frac{R_\odot}{r} \right)^{\ell+2} 
\left[ \frac{ 1 }{ \ell+1 } \right] 
\, \frac{\partial P_{\ell}^{0}( \cos{\theta} )}{\partial \theta} \,\,\, ,\\
B_{\phi}(r,\theta) &=& 0 \,\,\, ,
\end{eqnarray}
where $P_{\ell}^{m}(\cos{\theta})$ represents the corresponding $(\ell, m)$ Legendre polynomial and $R_\odot$ is the solar radius.

\begin{figure}[t]
    \centering
    \includegraphics[width=0.96\linewidth]{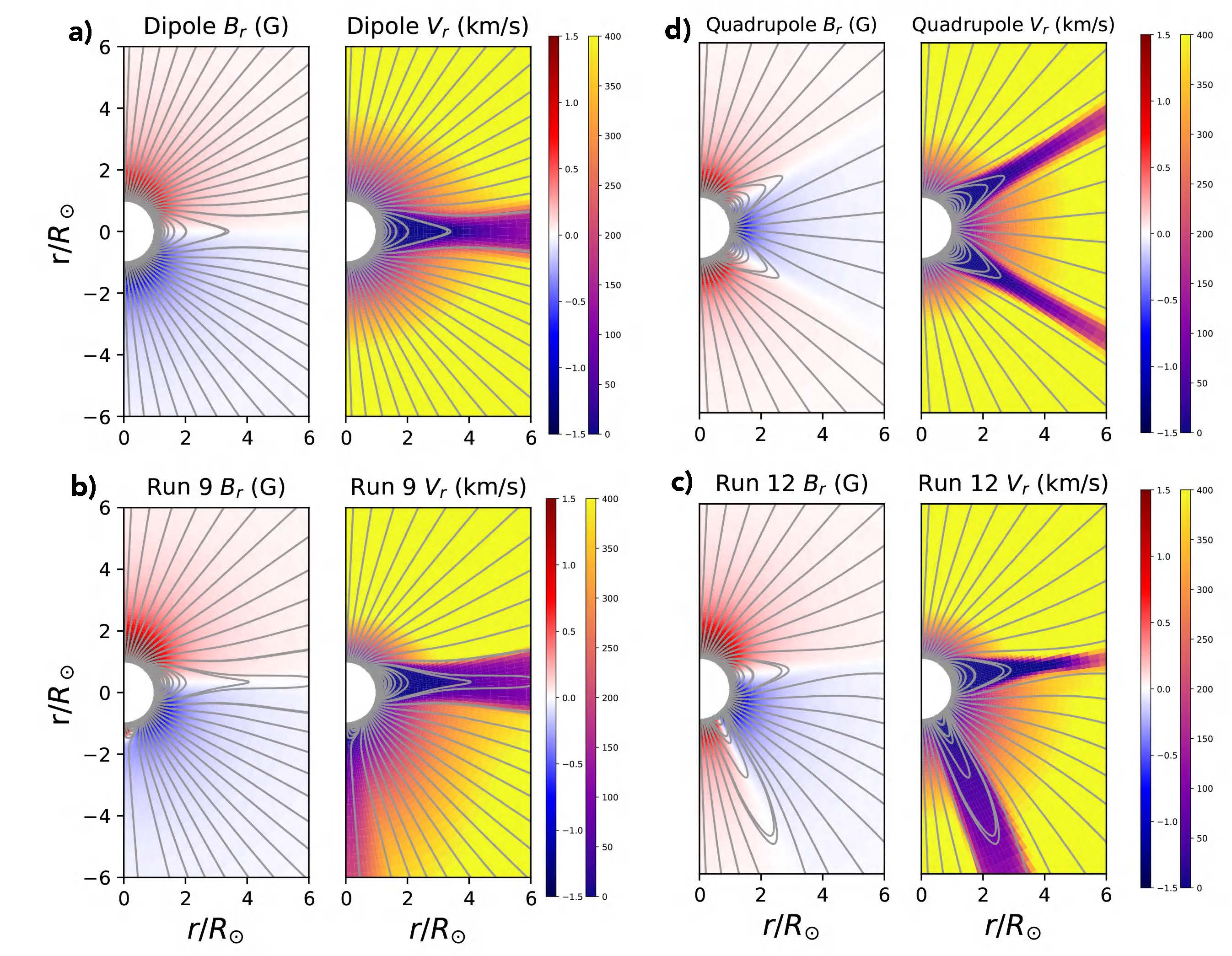}

    \caption{WindPredict-AW solar wind simulation results in four cases covering the range of relative dipole--quadrupole strengths: (a) a pure dipole (Run \#0: $g_{10}=8.2$, $g_{20}=0$ );  (b) a stable large-scale pseudostreamer (Run \#9, $g_{10}=8.2$, $g_{20}=7.38$ ); (c) an unstable pseudostreamer/secondary HCS case (Run \#12, $g_{10}=6.56$, $g_{20}=8.2$ ); and (d) a pure quadrupole (Run \#20: $g_{10}=0$, $g_{20}=8.2$). The HCSs are positioned above the closed-field helmet streamer loop systems corresponding to any $B_r = 0$ isosurfaces in the extended corona. An animated version of this figure is available as supplementary material.}
    \label{fig:dip/quad}
\end{figure}

We have performed 21 simulations spanning the $g_{10}$--$g_{20}$ parameter space with each coefficient magnitude ranging from $[0, 8.20]$~G in 10\% increments, and a few additional runs near the identified system behavior transitions. Run \#0 is therefore the pure dipole case $\left( g_{10}=8.2, \, g_{20}=0.0 \right)$ and Run \#20 corresponds to the pure quadrupole case $\left( g_{10}=0.0, \, g_{20}=8.2 \right)$. 
The relaxed solar wind solutions and global streamer structures for four representative cases are shown in Figure~\ref{fig:dip/quad} as panels (a)--(d), where we show $B_r(\mathbf{r})$ (left) and $V_r(\mathbf{r})$ (right) with magnetic field lines shown in gray. 
Figure~\ref{fig:dip/quad}(a) shows Run \#0, the pure dipole case, while Figure~\ref{fig:dip/quad}(d) shows Run \#20, the pure quadrupole case. Section~\ref{sec:results} will summarize the results of our parametric survey simulation runs, obtained by stepping systematically through the $(g_{10}, g_{20})$ parameter space.

The parameters that define the initial solar atmosphere, energy equation source terms, and Alf\'{e}nic wave-turbulence transport are determined as follows.
The gravitational constant and solar mass $M_\odot$ yield the gravitational acceleration $g_\odot = 2.74\times10^4$~cm~s$^{-2}$ at $r=R_\odot$. The base mass density is $\rho_\odot = 5.0 \times 10^{-16}$~g~cm$^{-3}$ corresponding to a proton number density $n_{p,\odot} \approx 3.0 \times10^{8}$~cm$^{-3}$.  The initial temperature at the base is taken as  $T_\odot = 0.5 \times 10^6$~K, consistent with the coronal transition region depicted in Figure 1b of Reville et al. (2018).
The parameters that determine the (ad hoc) background coronal heating contribution, $Q_h$, include the heating scale height, taken as $H=R_\odot$, and the photospheric energy flux $F_h = 2.0 \times 10^4$~erg~cm$^{-2}$~s$^{-1}$. 
The primary coronal heating mechanism however, is the wave-turbulence dissipation.
Wave heating parameter $\delta v_\odot$ essentially sets the average wave energy flux through lower boundary. Here we have taken $\delta v_{\odot} = 30.0$~km~s$^{-1}$ as in \citet{reville_2020b}. Given our average surface field magnitude $\langle B_{\odot} \rangle \approx 4.5$~G, the total input of  Alfv\'{e}nic wave energy is $\langle \rho_\odot \, \delta v_\odot ^2 \, v_A\rangle \approx 1.19 \times 10^ 5$~erg~cm$^{-2}$~s$^{-1}$. 
The wave-turbulence contribution to the plasma heating $Q^{\pm}_{\rm AW}$ has a characteristic dissipation scale $\lambda$ which is determined by the parameter $\lambda_{\odot} = 0.022 R_{\odot}$~G$^{1/2}$, corresponding to roughly the size of a supergranule \citep{Verdini2007}.
Thermal conduction parameters are given by the standard diffusion coefficient, $\kappa_0 = 9 \times 10^{-7}$~erg$^{-1}$~cm$^{-1}$~K$^{-1}$, and the characteristic distance of the heat flux transition to the collisionless regime, $r_{\rm coll} = 5 R_\odot$.   
Figure~\ref{fig:dip/quad} also shows the global structure of the solar wind outflow $v_r(r,\theta)$ as the right plot in panels \ref{fig:dip/quad}(a)--(d) for Runs \#0, \#9, \#12, and \#20, respectively. Each panel corresponds to simulation snapshots after $t=418.0$, $440.0$, $462.0$, and $431.2 $ hrs, respectively,  of solar wind relaxation in order to highlight the range of coronal and inner heliosphere topologies available---for the same model of solar wind---with just the first two ($\ell=1, 2$) terms of a spherical harmonic representation of the $t=0$ magnetic field.


\section{Results}
\label{sec:results}

%
%
%
%
%

\subsection{Global Structure of the Streamer Belt and HCS: Parametric Survey of Relative Dipole--Quadrupole Magnitude}

The four example simulation runs shown in Figure~\ref{fig:dip/quad} represent four distinct global magnetic field configurations resulting from the magnitudes of the spherical harmonic coefficients in our parameter space $g_{10},\, g_{20} \in [0,8.2]$~G.
This two-dimensional $(g_{10},g_{20})$ parameter space can be expressed in standard polar coordinates $( \varrho, \vartheta)$ defined with the usual 2D Cartesian-to-cylindrical coordinate transformation. For future convenience, we will refer to these space coordinates parameters as the magnitude $\varrho = M_{1,2}$ and as a normalized phase angle $\vartheta = \alpha (\pi/2)$. For our set of simulation runs, $\alpha \in [0,1]$ corresponding to $\vartheta \in [0,\pi/2]$, which yields the following relation 
\begin{equation}
M_{1,2} = \left[ \, g_{10}^2 + g_{20}^2 \, \right]^{1/2} \;\;,\;\;\;\;\; \alpha = \left( \frac{2}{\pi}\right) \sin^{-1}{\left[ \frac{g_{20}}{M_{1,2}} \right]} \;\; .
\end{equation}
%
%
Hereafter, we will use $\alpha$ as the single parameter to characterize different (angular) regions of the dipole--quadrupole $(g_{10},g_{20})$ coefficient space. 
The pure dipole case (Run \#0; Figure~\ref{fig:dip/quad}(a)), has $\alpha=0$ ($g_{20}=0$) and represents a Solar Minimum-like coronal configuration with a single equatorial closed-flux streamer belt extending into the HCS. 
The pure quadrupole case (Run \#20; Figure~\ref{fig:dip/quad}(d)) has $\alpha=1$ ($g_{10}=0$) and represents a more complex Solar Maximum-like structure with two distinct helmet streamer belts and two stable HCSs.
Figure~\ref{fig:dip/quad}(b) and \ref{fig:dip/quad}(c) show two intermediate, mixed cases each with $g_{10},\, g_{20} > 0$. 
Figure~\ref{fig:dip/quad}(b) shows the Run \#9 case $\alpha=0.476$, $\left( g_{10}=8.2, \, g_{20}=7.38 \right)$ which has a large \emph{stable} pseudostreamer in addition to a singular streamer belt and stable HCS.
Figure~\ref{fig:dip/quad}(c) shows the Run \#12 case $\alpha=0.570$, $\left( g_{10}=6.56, \, g_{20}=8.2 \right)$ which has a large \emph{unstable} pseudostreamer that develops the cyclic opening-up and closing-down of sufficient pseudostreamer flux to form a secondary ``polar crown'' helmet streamer belt with its associated conical HCS/HPS.
A number of our simulation runs show similar periodic partial openings of the PS flux. The stability (or longevity) of the secondary helmet streamers and HCSs/HPSs ultimately depends on the coronal flux distribution, represented here via the relative strengths of the $(g_{10}, \,g_{20})$ coefficients or their normalized phase angle $\alpha$. 

\begin{figure}[t]
    \centering
    \includegraphics[width=0.80\linewidth]{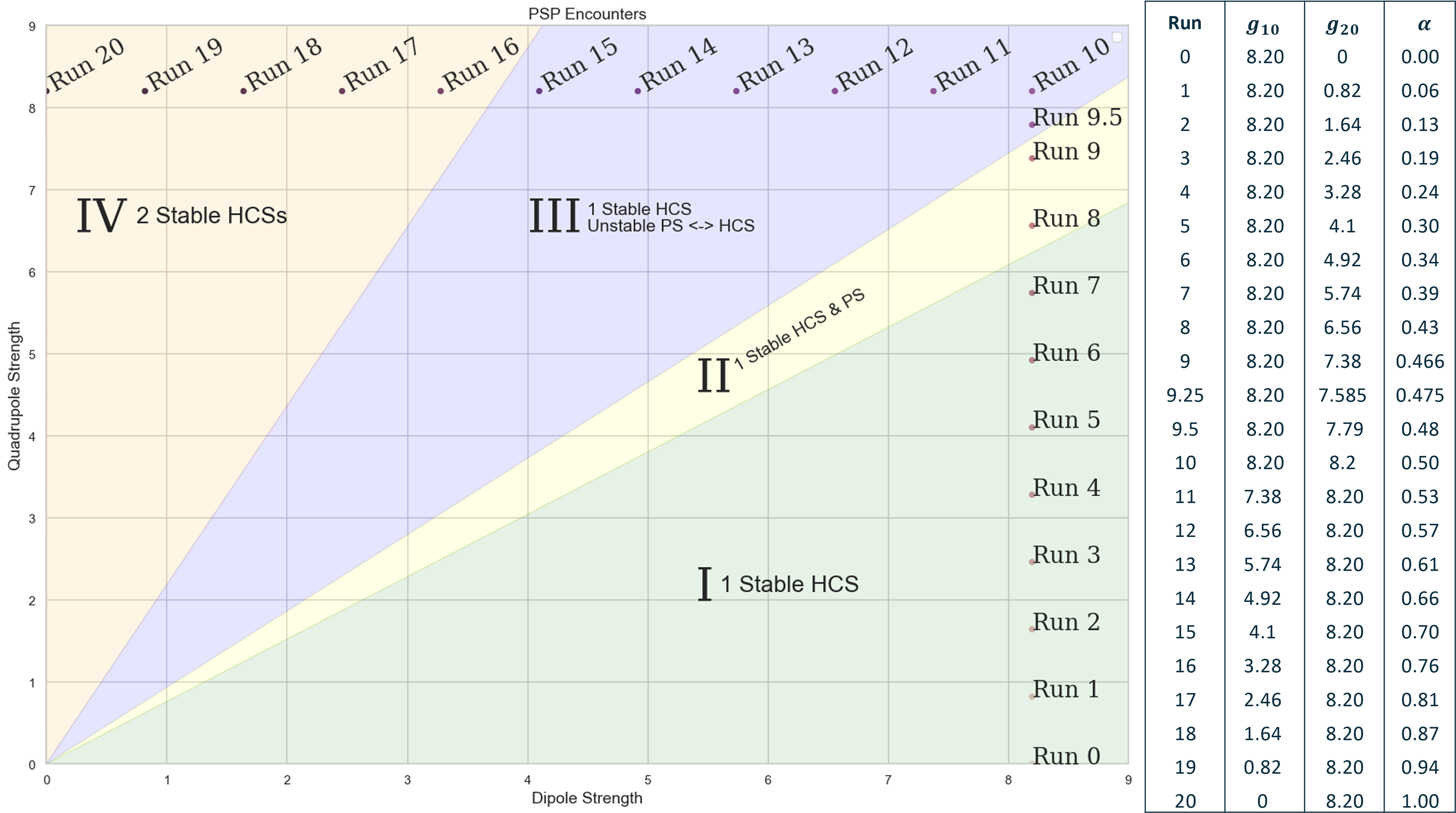}
    \caption{The set of simulations in our parametric survey of the $g_{10}$--$g_{20}$ dipole--quadrupole coefficient space for the coronal magnetic field. The shaded regions, I, II, III, and IV, represent distinct coronal--heliosphere topologies and resulting system dynamics.}
    \label{fig:param}
\end{figure}

In Figure~\ref{fig:param}, we display the full set of our 23 simulation runs by their locations in $(g_{10},g_{20})$ dipole--quadrupole coefficient space. 
The Figure~\ref{fig:param} legend shows the color and plot symbol assigned to each of our simulation Runs, the normalized phase angle $\alpha$, and the $(g_{10},g_{20})$ coefficient values (in G).
%
%
Here we have shaded the four distinct regions---defined by their angular extent (range of $\alpha$ values)---that correspond to different global-scale coronal magnetic field and plasma structure and/or different regimes of the system behavior and evolution. The four representative simulation snapshots shown in Figure~\ref{fig:dip/quad} illustrate some of the characteristic differences in the global-scale streamer flux system geometries and their corresponding HCS/HPS extension(s) that arise during the transition \emph{in parameter space} from dipole-dominated fields to quadrupole-dominated fields.
These four regimes are defined as follows.

\textbf{Region I.} The simulation runs in the range $\alpha \in [0,\, 0.430]$ (Runs \#0--\#7) all show a single and ``stable'' HCS location in a dipole-dominated, Solar minimum-like corona. These simulations contain only one global $B_r$ polarity inversion line (PIL) on the $r=R_\odot$ inner boundary under helmet streamer belt. Despite similar ``relaxed'' (quasi-steady) global-scale solar wind solutions, Runs \#0--\#7 exhibit systematic decreasing of the $|B_r|$ values at the South Pole as $g_{20}$ increases with $g_{10}$ fixed, i.e., as we start to introduce more of the quadrupole component. 

\textbf{Region II.} The simulations in the range $\alpha \in [0.430,\, 0.488]$ correspond to Runs \#8, \#9, and \#9.25. We note the fractional Run labels (Runs \#9.25, \#9.5) merely represent simulations with intermediate $g_{10}$, $g_{20}$ values between those of Runs \#9 and \#10. These cases are similar to Region~I in that they result in a quasi-stable HCS/HPS in a dipole-dominated, Solar minimum-like corona. The major topological differences between the I and II configurations results from the $g_{20}$ magnitude being large enough to change the sign of $B_r$ at the South Pole. This establishes a second global PIL at high-latitude and forms a new, \emph{stable} closed-flux region embedded within an open flux region of a single polarity---the precise topology of a coronal pseudostreamer ($\S$\ref{subsec:I_to_II}). 

\textbf{Region III.} The simulations in the range $\alpha \in [0.488,0.656]$ correspond to Runs \#9.5--\#13. Region III is the most dynamic and interesting region because these are the \emph{unstable} large-scale pseudostreamer cases. Here, the quadrupole magnitudes overtake the dipole magnitude $g_{20} \ge g_{10}$ but are not yet strong enough to maintain the quadrupole-dominated, stable two-HCS/HPS configuration. The SouthPole's positive radial flux---the amount of PS closed-flux in the simulation's $t=0$ potential field---is sufficient to create a large enough PS that the outer layers of the PS, including its topological separatrix boundary are dragged open by the ambient solar wind. However, this newly opened flux resulting from the narrow ``coronal funnel'' configuration does not stay open, after some time the whole secondary HCS/HPS structure is disrupted via a streamer-blowout type of eruptive transient from the polar crown helmet streamer (HS) which collapses back down to reform the PS flux system. Region III includes the runs that exhibit this cyclic, formation-and-collapse of their secondary conical HCSs during the opening-up (PS$\rightarrow$HS) and closing-down (HS$\rightarrow$PS) of the large-scale PS system in the southern coronal hole ($\S\S$\ref{subsec:II_to_III},~\ref{subsec:III}).

\textbf{Region IV.} The simulations in our final region, $\alpha \in [0.656,\, 1.0]$, correspond to Runs \#14--\#20, which are essentially the quadrupole version of Region I's Runs \#0--\#7. Here, each of these simulation runs form two stable HCSs/HPSs in a quadrupole-dominated, Solar Maximum-like corona. The radial magnetic flux distribution on the lower boundary continues to have two global PILs and therefore, two coronal helmet streamer belts. After the initial period of relaxation which creates the coronal/heliosphereic open flux regions and their HCS/HPS boundaries, these Runs never experience a catastrophic collapse (closing back down) of the secondary HCS/HPS structure and its open flux ($\S$\ref{subsec:III_to_IV}).

%
%

In the following sections, we will examine the transitions between the different topologies and system behavior of our four Regions (parameterized by $\alpha)$, in order to better understand the physical processes associated the dynamic evolution occurring during individual simulation runs and in our parametric survey as a whole.

\subsection{${\rm I}\rightarrow{\rm II}$. Formation of the Polar Pseuodstreamer}
\label{subsec:I_to_II}

 \begin{figure}[!t]
    \centering
    \includegraphics[width=0.85\linewidth]{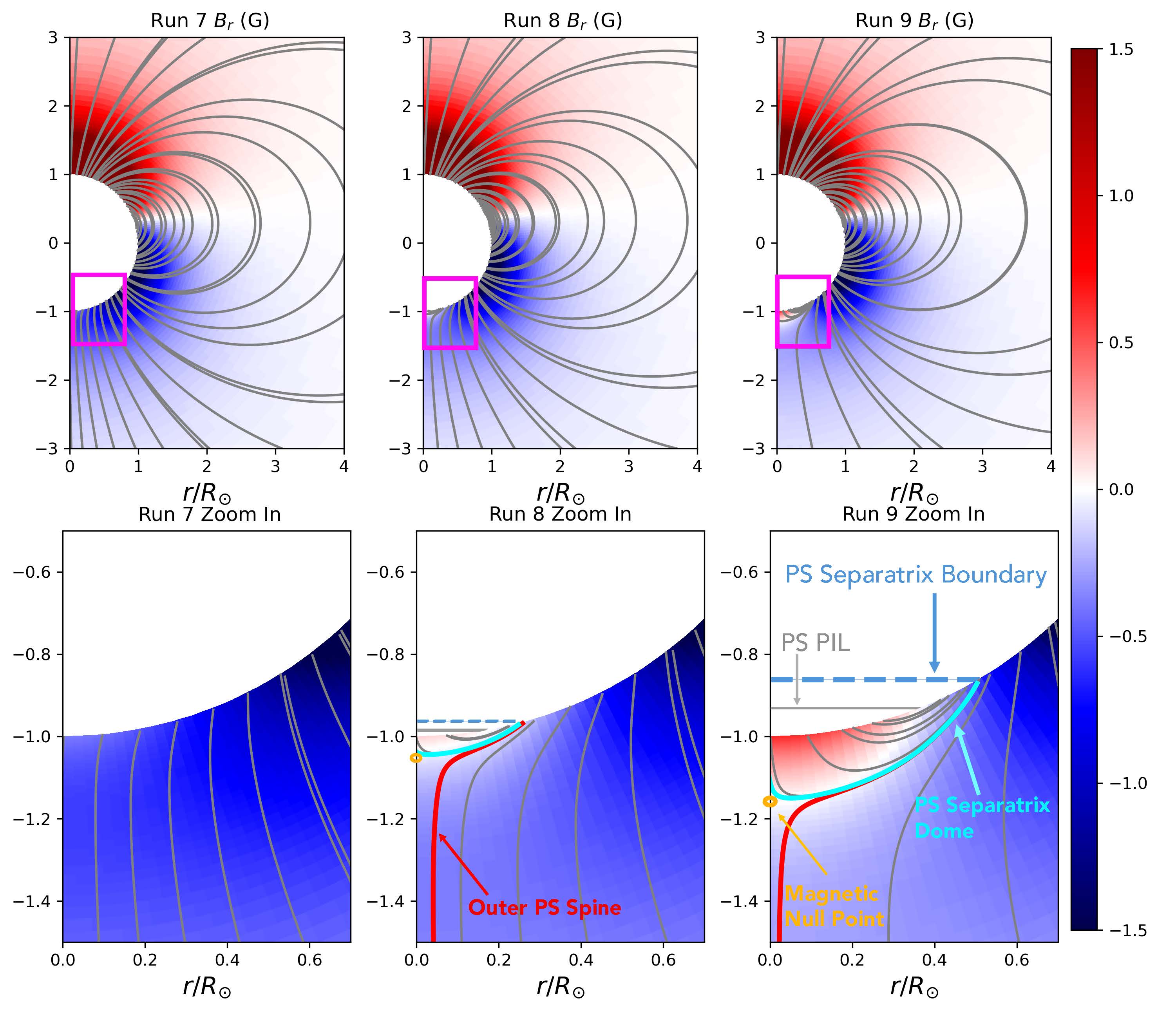}
    \caption{Magnetic field simulation results for Run 7 =$(g_{10}=8.2, g_{20}=5.74)$, run 8 =$(g_{10}8.2, g_{20}=6.56)$ and Run 9 = $(g_{10}=8.2, g_{20}=7.38)$ illustrating the emergence of a magnetic field opposite the field at the South Pole.}
    \label{fig:startevolution}
\end{figure}

The first major coronal topology transition in our simulation set---the creation of a coronal PS at the South Pole---occurs between Run \#7 and Run \#8, corresponding to a normalized phase angle of $\alpha_{\rm I\rightarrow II} = 0.405 \pm 0.015$ and shown in Figure~\ref{fig:param} as the dividing line between the green-shaded Region I and the yellow-shaded Region II.
Figure~\ref{fig:startevolution} illustrates the differences in the $t=0$ coronal magnetic field topology between Runs \#7--\#9 that arises with the gradual increase of the relative contribution of the quadrupole $g_{20}$ component. In each column we plot both the global inner corona $B_r(\mathbf{r})$ and an expanded view of the South Pole where the new PS closed-flux system forms.

The top row of Figure~\ref{fig:startevolution} indicates there is almost no qualitative difference in properties of the $B_r \sim 0$ minima that will be the location of the main (global) helmet-streamer belt and its associated HCS/HPS in Runs \#7--\#9.
In contrast, the bottom row of Figure~\ref{fig:startevolution} indicates there is a substantial qualitative difference in the radial flux distribution that eventually becomes the PS closed-flux.  
Figure~\ref{fig:startevolution}, middle column, shows the Run \#8 ($\alpha=0.430$) simulation where we observe a parasitic polarity area first appear at the South Pole. This area of opposite polarity facilitates the creation of a pseudostreamer.   The cyan line delineates the separatrix surface that separates the closed loops of the PS from the surrounding open field. The orange arrow marks the location of the 3D coronal null point of the PS flux system.
Figure~\ref{fig:startevolution}, right column, shows the Run \#9 ($\alpha=0.467$) results, which illustrates the continued      ``expansion'' of the PS closed-flux content as the area and magnitude of the parasitic polarity spot at the South Pole increases. The Region II simulations are stable PS configurations---the solar wind relaxation inflates the height and increases the mass density within the coronal PS but not to a great enough extent that they open up, even temporarily, without some additional energy source, such as low-lying, highly sheared or twisted magnetic field.

\subsection{${\rm II}\rightarrow{\rm III}$. Pseudostreamer Opening and Unstable Secondary HCS Formation}
\label{subsec:II_to_III}

%

The second major transition in both global-system behavior and coronal topology occurs at a normalized phase angle of $\alpha = 0.478 \pm 0.004$ between the simulation Runs \#9.25 and \#9.5 and represents the point at which the pseudostreamer flux system can no longer be considered \emph{stable} with respect to its interaction with and response to the ambient solar wind. 
Before this point ($\alpha \lesssim 0.478$) in Region II, the PS inflation during the MHD solar wind relaxation ultimately decreases with time and until the system achieves a quasi-steady equilibrium outflow. After this point ($\alpha \gtrsim 0.478$), the pseudostreamer inflation is able to open up enough of the outer layers that a run-away, positive feedback loop is established between the removal of restraining closed flux via opening/expansion into the solar wind and increased expansion of the PS flux in response \citep[see, e.g.,][for discussion of the helmet streamer equivalent]{Lynch2016b}.

\begin{figure}[!t]
    \centering
    \includegraphics[width=0.95\linewidth]{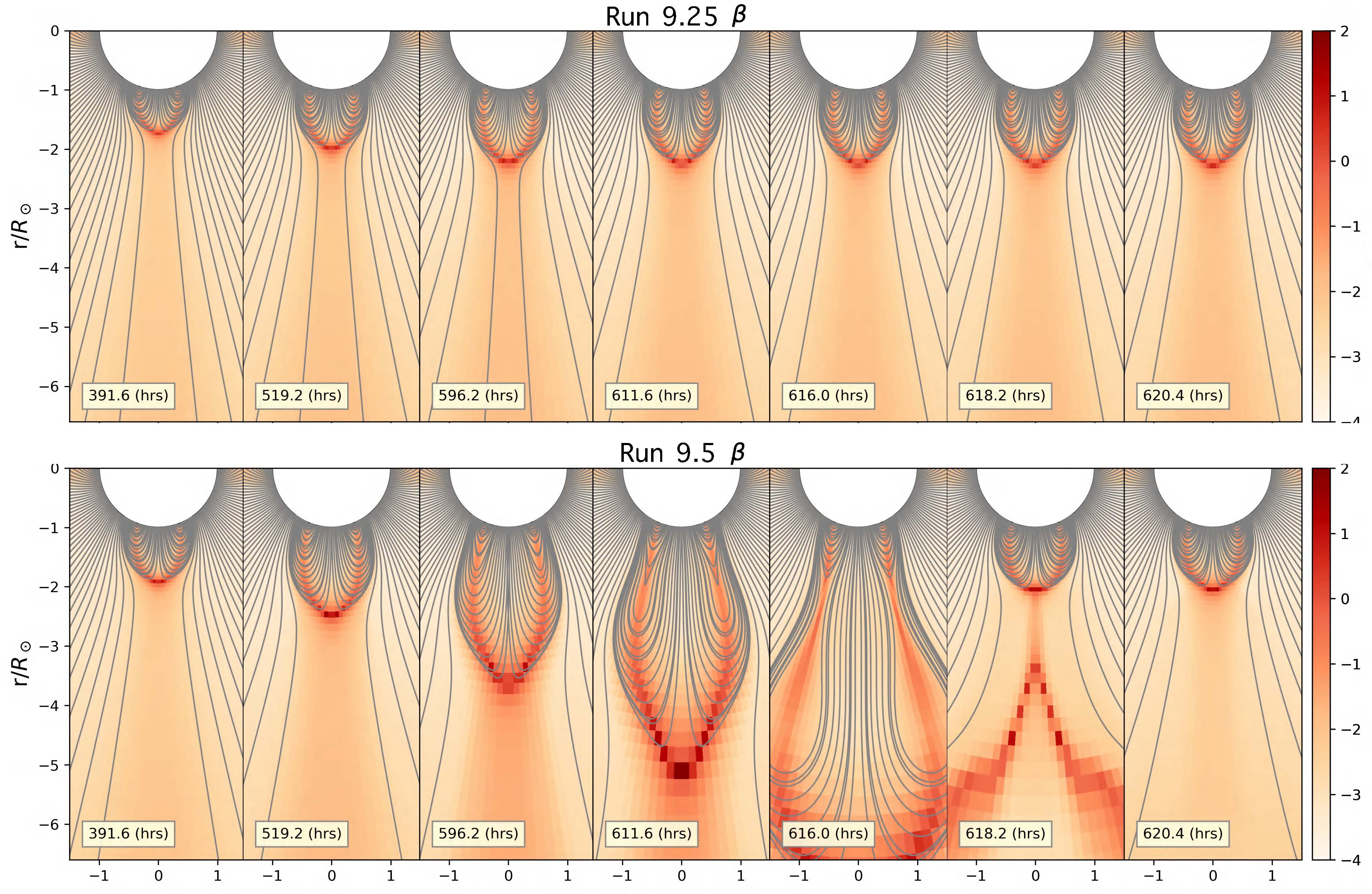}
    \caption{Comparison between PS evolution in the vicinity of the $\alpha = 0.478$ stability threshold. Representative magnetic field lines are shown in gray over a snapshot of the plasma $\beta$ ($\log_{10}$ scale). The top row shows the \emph{stable} PS of Run \#9.25 ($\alpha = 0.475$) and the bottom row shows \emph{unstable} PS of Run \#9.5 ($\alpha = 0.480$). See text for further details. An animated version of this figure is available as supplementary material.}
    \label{fig:comp1051075}
\end{figure}

Figure~\ref{fig:comp1051075} shows the comparison between the temporal evolution of the \emph{stable} pseudostreamer of Run \#9.25 ($\alpha = 0.475$; $g_{10}=8.2$, $g_{20}=7.585$), shown in the top row, and the \emph{unstable} pseudostreamer of Run \#9.5 ($\alpha = 0.480$; $g_{10}=8.2$, $g_{20}=7.79$), shown in the bottom row. 
The differences in the global system behavior resulting from an incremental 2.7\% increase in the $g_{20}$ coefficient magnitude is visually striking.
In Run \#9.25, the PS null-point and separatrix boundary reach a height of $\sim$2.25~$R_\odot$ and remains there for the duration of the simulation. In Run \#9.5 however, the PS null-point and separatrix boundary have already reached $\sim$2.5~$R_\odot$ by $t=519.2$~hrs---a height greater than the Run \#9.25 PS ever reaches. For our particular solar wind model, this height is close enough to the sonic critical point in the adjacent open-flux regions ($r_c \sim 1.5-2.5 R_\odot$) that the PS enters a regime of unstable, runaway expansion leading to the (temporary) opening of some small amount of parasitic polarity open flux in a ``coronal funnel'' configuration \citep{Panasenco2019}. This opening process is the transition of a large-scale coronal pseudostreamer into an isolated, secondary helmet streamer belt with its own HCS/HPS. 

This behavior provides a quantitative ``stability'' threshold for the ratio of the dipole to quadrupole magnetic field strengths required for a global-scale pseudostreamer to be susceptible to a significant nonlinear response and/or major disruptive transient in response to a small-scale perturbation. 
Once the PS has over-expanded, reconnection within the newly-formed conical HCS/HPS creates a disconnected/eruptive flux rope transient (a torus flux ring in our 2.5D axisymmetric system) that effectively ``blows out'' conical-streamer belt and accumulated current density structures, and collapses back down, reforming the large pseudostreamer in the wake of the eruption. We note the PS nullpoint and separatrix boundary height after the eruptive transient has returned to $\sim$2~R$_\odot$, suggesting the process may repeat. We find the process does repeat, and will examine this evolution below.

%

%

\subsection{${\rm III}$. Cyclic Unstable Secondary HCS Formation and Collapse}
\label{subsec:III}

We have defined Region~III in our $(g_{10}, \, g_{20})$ parameter space over the range $\alpha \in [0.478, \, 0.635]$ (the purple shaded region in Figure~\ref{fig:param}) which includes Runs \#9.5--\#13.
Each of these cases show the same global-scale cyclic evolution: ($i$) PS gradual expansion; ($ii$) runaway PS opening and transition into a localized helmet streamer--HCS/HPS formation; ($iii$) reconnection-generated streamer blowout-like transient; that leads to ($iv$) post-eruption secondary HCS/HPS collapse and PS reformation. In this section we present an analysis of the Run \#12 simulation results to characterize the coronal magnetic field and plasma signatures of these system dynamics.

\subsubsection{Open Flux Evolution}

\begin{figure}[t]
    \includegraphics[width=0.526\linewidth]{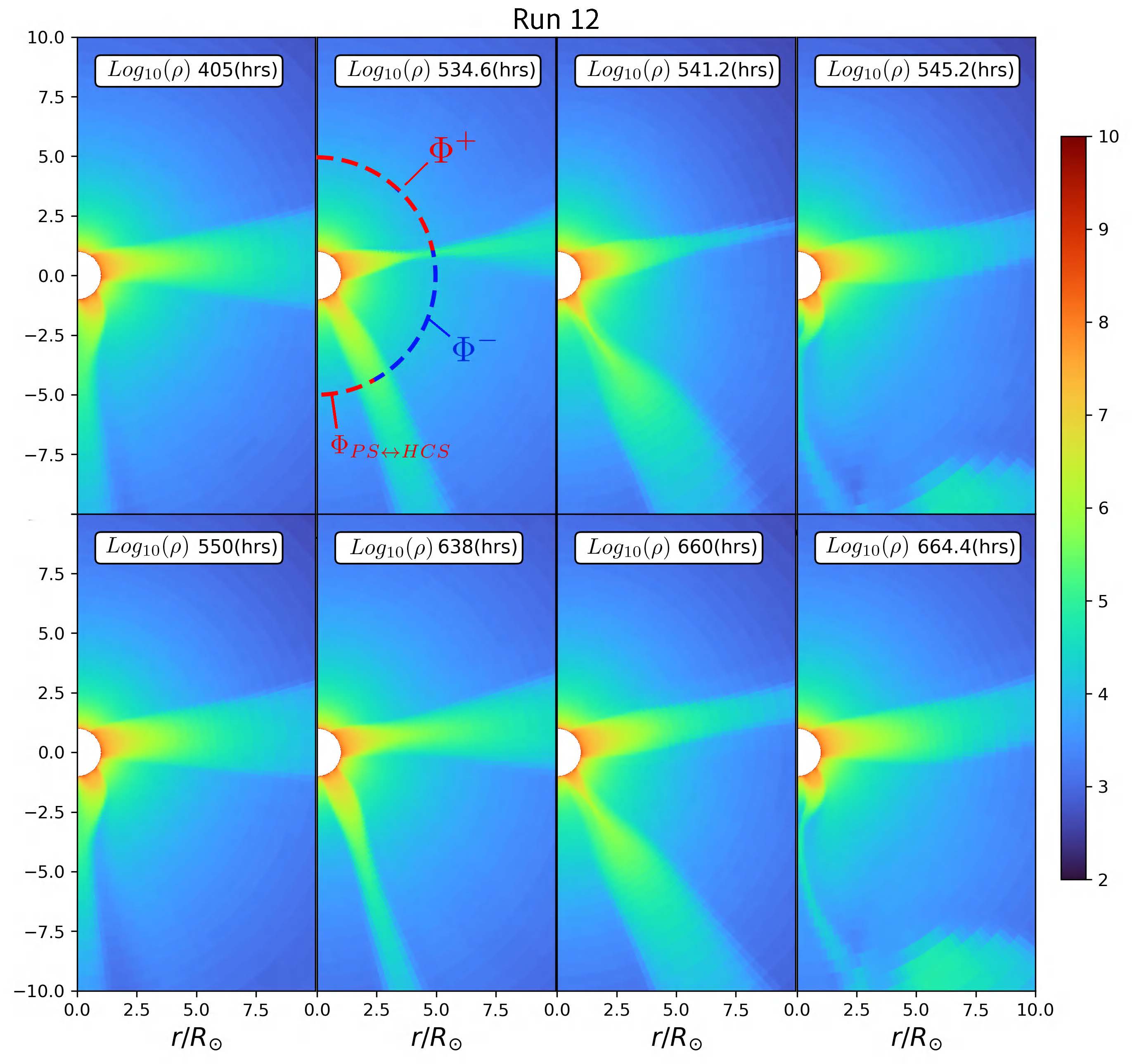}
    \includegraphics[width=0.474\linewidth]{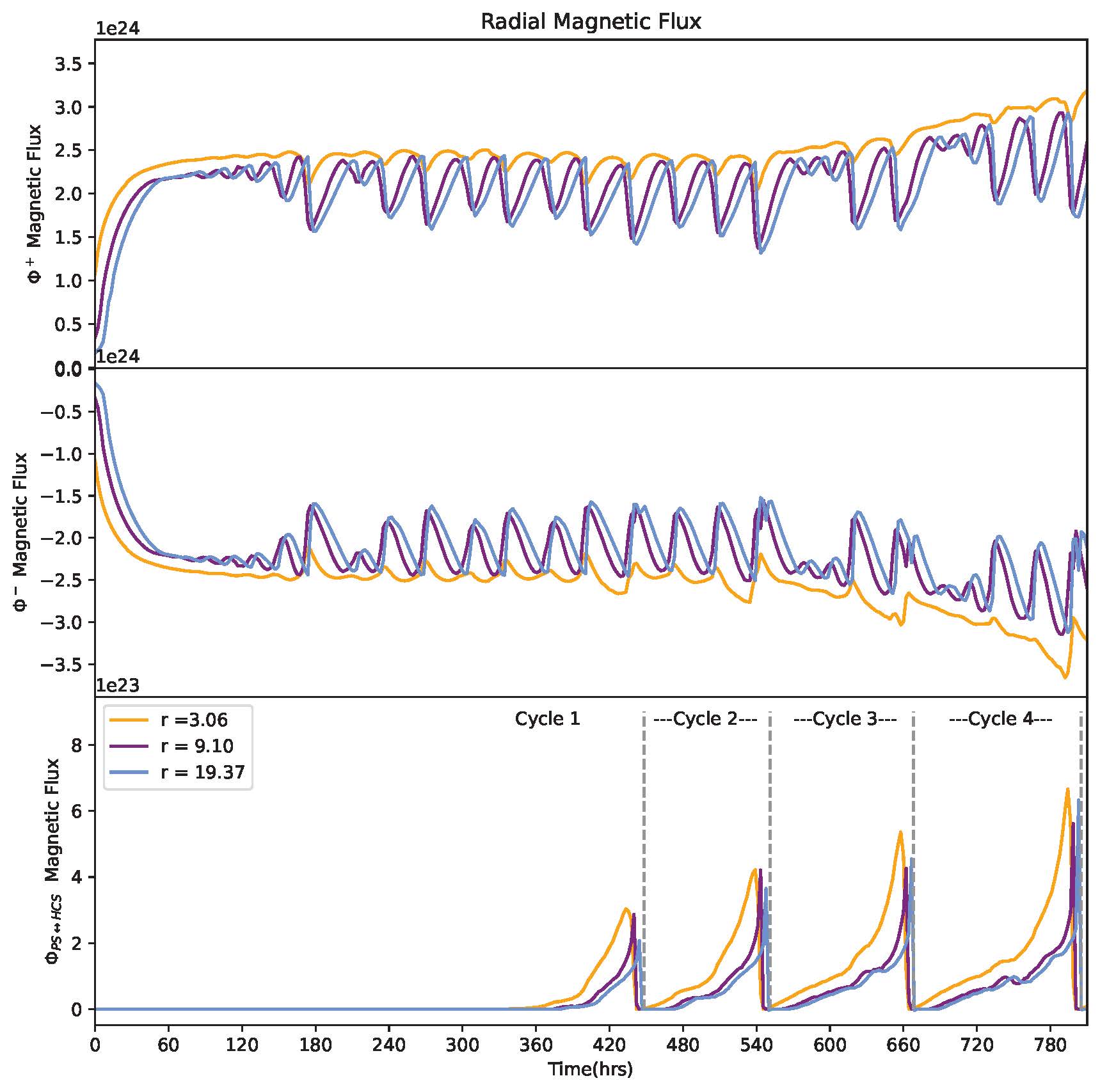}
    \caption{(a) Mass density evolution of simulation of Run 12, $\alpha=0.570$ $(g_{10}=6.56, \, g_{20}=8.2)$ during two cycles of the PS$\leftrightarrow$HS transition and secondary HCS formation and collapse. (b) The (signed) radial magnetic flux passing through the $R = [3.6R_\odot, 9.10R_\odot]$ surfaces for each of the open flux domains. The upper panel shows $\Phi^{+}$ (between the North Pole and the first HCS), the middle panel shows $\Phi^{-}$ (flux between the first HCS and either the South Pole or the unstable, secondary HCS), and the bottom panel shows $\Phi_{\rm PS\leftrightarrow HS}$ (between the second HCS and the South Pole, if applicable).}
    \label{fig:run8evol}
\end{figure}

The left column of Figure~\ref{fig:run8evol} illustrates the cyclic, global-scale evolution described above via snapshots of Run \#12 results for $\log_{10}$ of number density $n_p = (\rho / m_p)$. The top row shows the second pseudostreamer-to-helmet streamer (PS$\leftrightarrow$HS) cycle of opening up--streamer disruption--and closing back down ($t \in [405,\,545.3]$~hr). The bottom row shows the corresponding snapshots from the third PS$\leftrightarrow$HS cycle ($t \in [550,\,664.4]$~hr). 
The right column of Figure~\ref{fig:run8evol} quantifies the time-dependence of the open radial magnetic flux associated with each of our topological domains during the extended Run \#12 simulation, capturing four full PS$\leftrightarrow$HS cycles that form and disrupt/blow-out the secondary HCS/HPS.


The (signed) open magnetic flux is calculated as the standard $\Phi = \int \mathbf{B} \cdot d\mathbf{A}$ integral where the surface is a sphere at constant radius $R$ (i.e., $d\mathbf{A} = R^2 \sin{\theta}\, d\theta \, d\phi \, \mathbf{\hat{r}}$), usually taken beyond the Aflv\'{e}n surface where the closing down of flux back to the corona is no longer possible. In order to keep track of the signed radial flux in each of the open flux regions, we separate the surface integral into latitudinal ranges defined by the location of the HCSs (equivalently the $B_r = 0$ contours on the sphere at $r=R$). Given the 2.5D axisymmetry of our system, if the main streamer belt HCS is at a latitude $\Theta_{\rm HCS1}$, then the positive $(+)$ open radial flux in the northern is given by the $\theta$-integration limits $\theta \in [0, \Theta_{\rm HCS1}]$.
The limits of the $\theta$-integration for the negative $(-)$ open radial flux, $\theta \in [\Theta_{\rm HCS1}, \Theta_2]$ depend on whether or not the southern PS has opened up into the full secondary helmet streamer belt encircling the southern pole. The $\Theta_{2}$ limit when the PS is closed is $\Theta_{2} = \pi$. This occurs for every simulation run in Regions I and II, and for some simulation times during runs in Region III. For the Region III simulation times when the PS has opened up, the negative open flux is calculated until the position of the secondary conical HCS at $\Theta_2 = \Theta_{\rm HCS2}$.
In the simulations with two HCSs, the third radial flux region corresponds to the remaining $(+)$ flux over the SouthPole in either the stable Region IV runs or the unstable, temporary openings in the Region III simulations. Therefore, for a simulation output file at time $t$, the exact HCS $\theta$-positions are also time-dependent, leading to the following expressions for the instantaneous open fluxes:
\begin{eqnarray}
        \Phi^{+}(t) &=& 2\pi R^2  \int_{0}^{\Theta_{\rm HCS1}(t)} \, B_r(R,\theta,t) \, \sin{\theta} d\theta \, , \\
        \Phi^{-}(t) &=& 2\pi R^2  \int_{\Theta_{\rm HCS1}(t)}^{\Theta_2(t)} \, B_r(R,\theta,t) \, \sin{\theta} d\theta \, , \\
         \Phi_{\rm PS\leftrightarrow HS}(t) &=& 2\pi R^2  \int_{\Theta_2(t)}^\pi \, B_r(R,\theta,t) \, \sin{\theta} d\theta \, ,
\end{eqnarray}
where we note that $\Phi_{\rm PS\leftrightarrow HS}$, corresponding to the previously-closed PS flux that is now ``open'' due to passing through $R$, is identically zero in the simulation output times with only one HCS.
The right column of Figure~\ref{fig:run8evol} plots the radial open flux quantities, from top to bottom, $\Phi^{+}(t)$, $\Phi^{-}(t)$, and $\Phi_{\rm PS\leftrightarrow HS}(t)$ for the Run \#12 simulation at the radial heights, $R=[3.06, \, 9.10, \, 19.7]$~R$_\odot$. The signatures of the two major time-dependent cyclic or quasi-periodic processes associated with our ``quasi-steady'' solar wind solutions are immediately obvious in the open flux quantities.

The $\Phi^{+}$ and $\Phi^{-}$ time series (top, middle panels) illustrate the well-known signatures of the quasi-periodic formation and release of magnetic island plasmoids from the cusp of the helmet streamer belt and the base of HSC1 \citep{higginson_lynch_2018,Lynch2020,Reville2020ApJL,reville_flux_2022}. These plasmoids correspond to the well-known ``streamer blob'' features identified in coronagraph imagery and are a relatively well-understood phenomena \citep{Sheeley1997,Sheeley2007,Rouillard2010a,Rouillard2010b,Viall2015,kepko2016}.
For $t \in [ \, 127,\, 800\, ]$~hr, there are 23 local maxima/minima (peaks) in the stable HCS1-associated open fluxes $\Phi^{+}$, $\Phi^{-}$ yielding 22 periods in the extended simulation coverage. This yields a mean streamer blob ejection period of $\langle \tau \rangle = 30.5 \pm 6.6$~hrs, which is fairly close to the periods obtained by \citet{Reville2020ApJL} between bursts of non-linear tearing onset and rapid growth in the eruption of plasmoids in the HCS of their dipolar, Solar Minimum-like wind simulation.


The bottom panel shows the magnetic flux from the lower HCS to the SouthPole $\Phi_{{\rm PS}\leftrightarrow{\rm HS}}$.
We calculate the periods between PS opening and secondary HCS/HPS collapse from the vertical lines denoting the duration of cycles 2--4 to obtain $\tau_{{\rm PS}\leftrightarrow{\rm HS}} = \left[ \, 4.29, \, 4.96, \, 5.67 \, \right]$~days, which yield a mean of  $\langle \tau_{{\rm PS}\leftrightarrow{\rm HS}} \rangle = 4.97 \pm 0.69$~days. 
We note the period is slowly increasing and the interaction between secondary HCS/HPS formation and disruption starts to influence the recurrent streamer blob production in the stable, Northern hemisphere HCS1. This can be seen in the $\Phi^{+}$, $\Phi^{-}$ plots right after the catastrophic collapse of the secondary HCS/HPS sheets after cycles 2 and 3. The helmet streamer blob size and flux content (and therefore the variation in our instantaneous open flux measure) begin to be affected by the global-scale pseudostreamer evolutionary cycle.
 

\subsubsection{Plasma Dynamics and Eruptive Transients}

\begin{figure}[!t]
    \centering
    \includegraphics[width=1.0\linewidth]{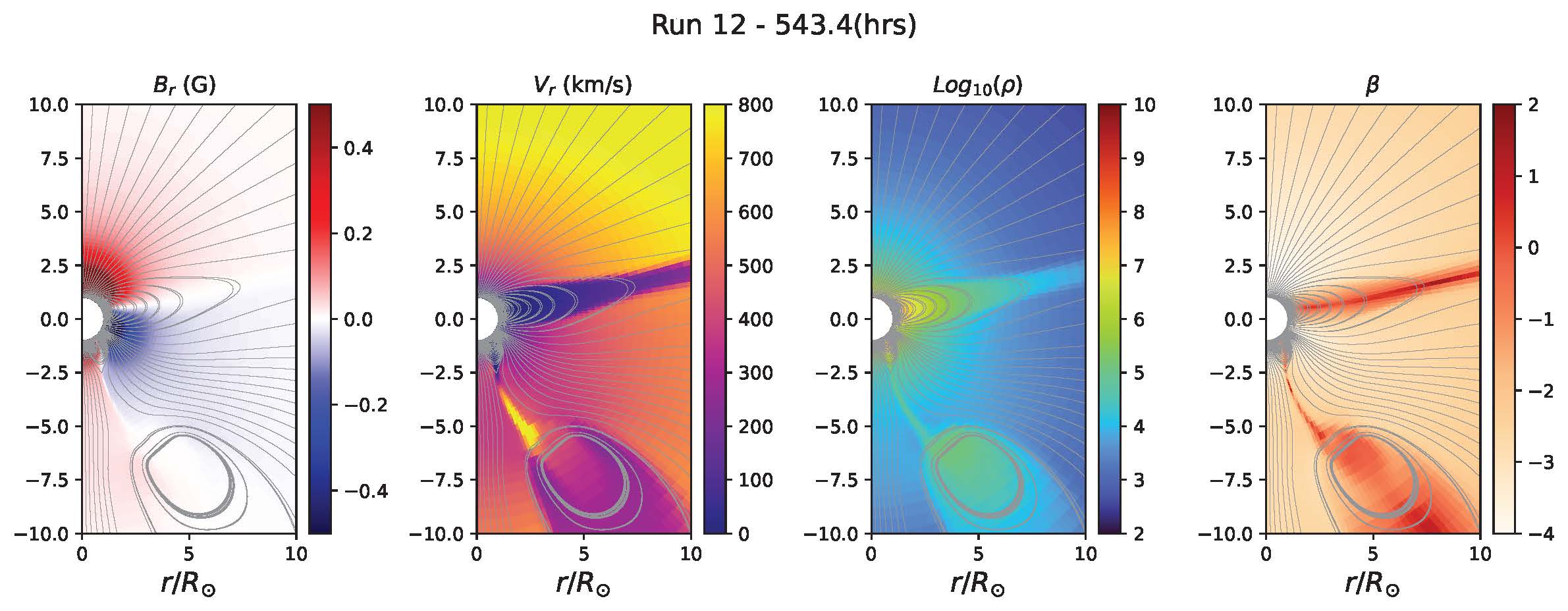}
    \caption{Snapshot of plasma and field quantities during the simulation Run \#12 ($\alpha=0.57$; $g_{10}=8.2$, $g_{20}=6.56$) eruption of a ``streamer blowout'' type of CME-like transient that disrupts the secondary HCS/HPS and causes the whole newly opened open flux region and its corresponding solar wind stream to collapse in the wake of the eruptive transient and reform the large-scale polar pseudostreamer closed flux system. From left to right we show $B_r$, $V_r$, $\log_{10}[\, n_p\,]$, and $\log_{10}[\, \beta \,]$. An animated version of this figure is available as supplementary material.}
    \label{fig:run12}
\end{figure}

Figure~\ref{fig:run12} shows four snapshot of plasma and field quantities at $t=543.4$~hr during the simulation Run \#12 ($\alpha=0.57$; $g_{10}=8.2$, $g_{20}=6.56$) eruption of a ``streamer blowout'' type of CME-like transient that disrupts the secondary HCS/HPS. This CME-like transient causes the entire newly-opened open-flux region and its corresponding solar wind stream to collapse in its wake. This collapse is facilitated via the same magnetic reconnection that forms the disconnected flux rope streamer transient analogous to the classic 2D CME initiation scenario for flare current sheet reconnection and CME eruptions \citep[][and references therein]{Linker2003,Janvier2015,Lynch2016b}. 
From left-to-right, Figure~\ref{fig:run12} plots $B_r(\mathbf{r})$, $V_r(\mathbf{r})$, $\log_{10}$ of number density ($n_p=\rho/m_p$) and $\log_{10}$ of plasma $\beta$. Representative magnetic field lines are over-plotted in each panel and show the large, disconnected plasmoid flux-rope transient. The fate of the previously closed PS flux that now opens can be seen from the geometry of the streamer blowout transient. The solar flare-like reconnection (in the portion of the HCS/HPS below the reconnection-generated flux rope transient) is transferring some portion of the red polarity open flux and blue polarity open flux into the erupting structure and back into the closed secondary HS belt flux system. Eventually all of the red polarity open flux is reconnected as the high-latitude closed flux system rebuilds, eventually reforming the large-scale polar pseudostreamer structure below a height of $\sim$2.5$R_\odot$ (e.g. Figure~\ref{fig:comp1051075}).

\subsection{${\rm III}\rightarrow{\rm IV}$. Transition to a Stable Multi-HCS/HPS Corona and Inner Heliosphere}
\label{subsec:III_to_IV}

The final transition in our system is between Runs \#13 and \#14 corresponding to a normalized phase angle of $\alpha = 0.635 \pm 0.025$. This transition represents going from the \emph{unstable} PS$\leftrightarrow$HS Region III configurations with cyclic formation and collapse of the secondary HCS/HPS structures to the \emph{stable}, quadrupole-dominated, two HCS/HPS configurations in Region IV.
The Region III Runs \#11--\#13 have average PS opening periods of $\tau_{{\rm PS}\leftrightarrow{\rm HS}} = [\, 5.31\pm 0.50,\, 4.62\pm0.47,\, 4.04\pm0.24\,]$~days, respectively.
%
%
For each of the Region IV simulations, the solar wind relaxation dynamics include only the initial opening of the $t=0$ closed fields with the establishment of the ambient solar wind outflow. Every simulation run includes this phase, e.g., see the $\Phi^{\pm}(t \lesssim 60~{\rm hr})$ behavior in Figure~\ref{fig:run8evol}, but for Runs \#14--\#20, once this initial opening creates the three separate open flux domains, it is clear the quadrupole component dominates the global flux distribution: the quasi-steady state closed flux content of the southern hemisphere HS belt, the location of its PIL on the lower boundary, and the resulting secondary HCS/HPS sheet extension of the streamer belt all systematically approach the equatorial symmetry of the pure quadrupole corona ($g_{10}=0$) of Figure~\ref{fig:dip/quad}(d).  
%


\section{Summary and Discussion \label{sec:disc}}


We have performed a parametric survey of idealized, axisymmetric solar wind MHD simulations in which the global coronal magnetic field configuration is varied through the dipole ($g_{10}$) and quadrupole ($g_{20}$) spherical harmonic coefficient magnitudes. 
We have identified four distinct regimes of system behavior resulting from the global magnetic field configuration on the basis of a normalized phase angle $\alpha \in [0,\, 1]$ between the $g_{10}$, $ g_{20}$ components.  
Our Regions I simulation Runs \#0--\#7 ($\alpha < 0.40$) and Region IV Runs \#14--\#20 ($\alpha > 0.65$) show a time-stationary dipole-dominated or quadrupole-dominated coronal structure with one or two HS belts and HCSs/HPSs, respectively.
In Region II, the $g_{20}$ value was sufficient to create a parasitic polarity flux concentration at the SouthPole, giving rise to a \emph{stable} large-scale coronal pseudostreamer flux system that remains closed. 
In Region III ($0.48 < \alpha < 0.65$), the relative quadrupole contribution has increased enough such that the closed PS flux system can be considered \emph{unstable} on the timescale of days-to-weeks and shows a cyclic process of opening up and forming a new, secondary HCS/HPS surrounding the coronal funnel solar wind stream rooted in the center of the parasitic polarity spot, followed by reconnection onset, rapid collapse of the secondary HCS in the wake of a streamer blowout-like transient disruption, and the reformation of the large-scale PS. 
Our results demonstrate that as the \emph{relative} contribution of the dipole component weakens, the increasingly dominant contribution of the quadrupole component drives the system toward the formation of new open flux domains with the associated secondary streamer belt and HCS/HPS extension required of the magnetic topology.

\begin{figure}[t]
    \centering
    \includegraphics[width=0.99\linewidth]{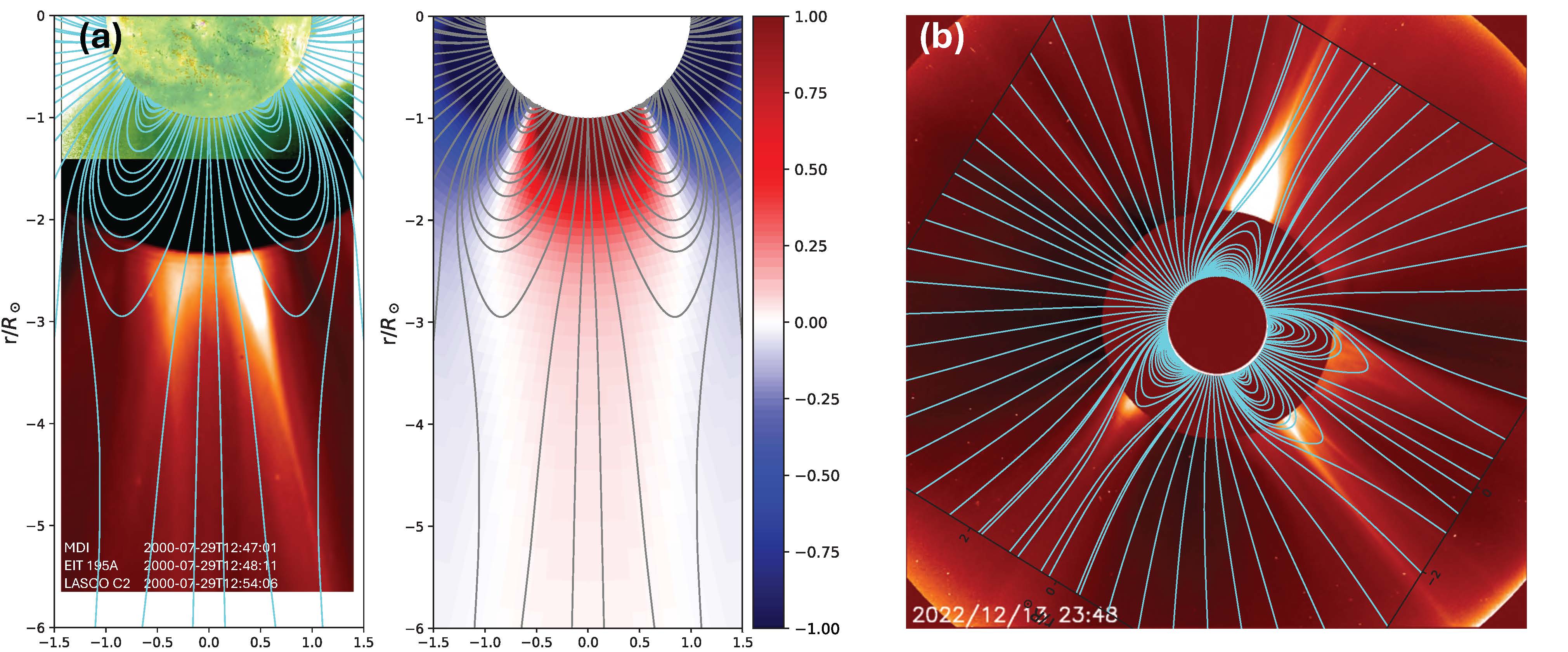}
    \caption{Qualitative comparison between two observed ``polar helmet streamer'' configurations and our idealized MHD simulations.
    (a) Left panel shows 29-Jul-2000 MDI, EIT 195\AA\ and LASCO/C2 observations of a polar helmet streamer belt over the North solarPole, adapted from \citet{Zhukov2008} and inverted here, along with the Run \#12 simulation's magnetic field lines at $t=389.4$~hr. Right panel shows the same field lines over $B_r(\mathbf{r})$ to illustrate the open flux polarities.
    (b) LASCO/C2 observation of the white-light corona on 13-Dec-2022 at 23:48UT compared with the same Run \#12 field lines shown in (a), now rotated by 58$^\circ$ counter-clockwise.}
    \label{fig:obs}
\end{figure}


Given the extremely idealized nature of axisymmetric modeling, it is worth a brief qualitative comparison between the spatial extent of our global-scale PS$\leftrightarrow$HS transition and the resulting secondary HS belt. 
The left panel of Figure~\ref{fig:obs}(a), adapted from an analysis of polar helmet streamers by \citet{Zhukov2008}, shows a composite of SOHO MDI, EIT 195\AA\, and LASCO C2 data for 29-Jul-2000 when a high-latitude filament channel (and its underlying photospheric $B_r = 0$ PIL) surrounded the entire NorthPole. Here we have inverted the original orientation of the observations for ease of comparison with our simulation geometry.
This secondary, closed HS belt as a ``double streamer'' configuration in the coronagraph plane-of-the-sky cross-section of Thomson-scattered white light intensity \citep{Rachmeler2014}.
The right panel of Figure~\ref{fig:obs}(a) shows the same Run \#12 ($\alpha=0.57$), $t=389.4$~hr simulation magnetic field line snapshot, now plotted over the simulation $B_r(r,\theta,t)$ distribution shown in the red--blue colorscale to highlight the $(+)$ polarity of the newly opened flux region surrounded by $(-)$ polarity open fields. 
In Figure~\ref{fig:obs}(b), we show a LASCO C2 white-light observation on 13-Dec-2022 at 23:48UT, alongside the same simulation magnetic field line snapshot in panel (a), but now over a larger portion of the domain including the main streamer belt, and rotated 58$^{\circ}$ counter-clockwise. 
The overall placement of both the main helmet streamer belt and the secondary PS$\leftrightarrow$HS belt are more favorable than one may have guessed from the model's magnetic field simplicity. Both the apparent conical opening angle and the width of the polar streamers in panel \ref{fig:obs}(a) appear larger than the observations, but the general agreement seems to suggest that, at least some of the time, the global magnetic structure of the extended corona and inner heliosphere is reasonably well-represented by the interplay between the first two terms of a spherical harmonic representation of the observed photospheric fields and their coronal extrapolations.   



The solar-cycle dependence of the lowest order multipoles of PFSS extrapolations of the photospheric magnetic field observations has been well-established \citep{Wang2014SSR}. Therefore, throughout the paper, we have referred to the pure dipole ($\alpha=0$) and pure quadrupole ($\alpha=1$) simulation cases as a Solar Minimum-like and a Solar Maximum-like global coronal and heliospheric structure, respectively. 
It is worth pointing out that the two examples of observed near-axisymmetric polar streamer configurations in Figure~\ref{fig:obs} are $\sim$22~years apart, with July 2000 being during Solar Maximum of Cycle 23 and December 2022 being during Solar Maximum of Cycle 25.
While we have described the transversal through our coefficient parameter space as starting from a dipole and increasing the quadrupole contribution (approximating an activity-cycle ascending phase), the axisymmetry precludes us from explicitly modeling the non-axisymmetric component of the dipole term ($|m|=1$), which is required to obtain a fully 3D representation of a ``tilted'' dipole configuration, i.e. as in Figure~\ref{fig:obs}(b). The polar field reversal however, is not due to the ``emergence'' of new cycle polar flux, rather the poleward ``convergence'' of new cycle flux introduced at lower latitudes on small scales as bipolar active region concentrations that evolve under the influence of differential rotation, meridional flow, and granular diffusion \citep{Caplan2025} to become weaker and more dispersed/larger-scale photospheric flux distributions that spread poleward to ultimately cancel the remaining old-cycle polar flux \citep[e.g.,][]{WangYM2003,Hathaway2015}. Had we described our parameter space traversal starting from the pure quadrupole case and increasing the relative contribution of the dipole component, this would correspond to an activity cycle descending phase with the dipole component representing the new cycle's polar field orientation.    

Our results also suggest there may be quantitative thresholds developed from observations---e.g., an $\alpha$-like analog---that would not only ``predict'' the potential for a more complex heliospheric sector structure on Carrington Rotation timescales, but could also help to determine when a global-scale PS configuration could be on the verge of a significant disruption. 
Both of these research avenues are being actively pursued via further numerical modeling and multi-viewpoint and multi-spacecraft observations and analyses of these global-scale coronal and heliospheric evolutionary processes.




\begin{acknowledgments}
The authors acknowledge support from NASA GSFC (80NSSC26K0467), XRP (80NSSC25K7545), and LWS (80NSSC24K1121). V.R. acknowledges support from CNRS and CNES through the APR program. 
\end{acknowledgments}


\bibliographystyle{aasjournal}
\bibliography{sample7}

@article{Wang2014SSR,
	adsurl = {https://ui.adsabs.harvard.edu/abs/2014SSRv..186..387W},
	author = {{Wang}, Y.-M.},
	doi = {10.1007/s11214-014-0051-9},
	journal = {\ssr},
	month = dec,
	number = {1-4},
	pages = {387-407},
	title = {{Solar Cycle Variation of the Sun's Low-Order Magnetic Multipoles: Heliospheric Consequences}},
	volume = {186},
	year = 2014}

@article{WangYM2003,
	adsurl = {https://ui.adsabs.harvard.edu/abs/2003ApJ...599.1404W},
	author = {{Wang}, Y.-M. and {Sheeley}, Jr., N.~R.},
	doi = {10.1086/379348},
	journal = {\apj},
	month = dec,
	number = {2},
	pages = {1404-1417},
	title = {{On the Topological Evolution of the Coronal Magnetic Field During the Solar Cycle}},
	volume = {599},
	year = 2003}

@article{Kepko2016,
	adsurl = {http://adsabs.harvard.edu/abs/2016GeoRL..43.4089K},
	author = {{Kepko}, L. and {Viall}, N.~M. and {Antiochos}, S.~K. and {Lepri}, S.~T. and {Kasper}, J.~C. and {Weberg}, M.},
	doi = {10.1002/2016GL068607},
	journal = {\grl},
	month = may,
	pages = {4089-4097},
	title = {{Implications of L1 observations for slow solar wind formation by solar reconnection}},
	volume = 43,
	year = 2016}

@article{Luhmann2020,
	adsurl = {https://ui.adsabs.harvard.edu/abs/2020SoPh..295...61L},
	author = {{Luhmann}, J.~G. and {Gopalswamy}, N. and {Jian}, L.~K. and {Lugaz}, N.},
	doi = {10.1007/s11207-020-01624-0},
	eid = {61},
	journal = {\solphys},
	month = apr,
	number = {4},
	pages = {61},
	title = {{ICME Evolution in the Inner Heliosphere}},
	volume = {295},
	year = 2020}

@article{Manchester2017,
	adsurl = {http://adsabs.harvard.edu/abs/2017SSRv..212.1159M},
	author = {{Manchester}, W. and {Kilpua}, E.~K.~J. and {Liu}, Y.~D. and {Lugaz}, N. and {Riley}, P. and {T{\"o}r{\"o}k}, T. and {Vr{\v s}nak}, B.},
	doi = {10.1007/s11214-017-0394-0},
	journal = {\ssr},
	month = nov,
	pages = {1159-1219},
	title = {{The Physical Processes of CME/ICME Evolution}},
	volume = 212,
	year = 2017
	}

@article{Abbo2016,
	adsurl = {https://ui.adsabs.harvard.edu/abs/2016SSRv..201...55A},
	author = {{Abbo}, L. and {Ofman}, L. and {Antiochos}, S.~K. and {Hansteen}, V.~H. and {Harra}, L. and {Ko}, Y. -K. and {Lapenta}, G. and {Li}, B. and {Riley}, P. and {Strachan}, L. and {von Steiger}, R. and {Wang}, Y. -M.},
	doi = {10.1007/s11214-016-0264-1},
	journal = {\ssr},
	month = nov,
	number = {1-4},
	pages = {55-108},
	title = {{Slow Solar Wind: Observations and Modeling}},
	volume = {201},
	year = 2016}

@article{Fox2016,
	adsurl = {https://ui.adsabs.harvard.edu/abs/2016SSRv..204....7F},
	author = {{Fox}, N.~J. and {Velli}, M.~C. and {Bale}, S.~D. and {Decker}, R. and {Driesman}, A. and {Howard}, R.~A. and {Kasper}, J.~C. and {Kinnison}, J. and {Kusterer}, M. and {Lario}, D. and {Lockwood}, M.~K. and {McComas}, D.~J. and {Raouafi}, N.~E. and {Szabo}, A.},
	doi = {10.1007/s11214-015-0211-6},
	journal = {\ssr},
	month = {Dec},
	number = {1-4},
	pages = {7-48},
	title = {{The Solar Probe Plus Mission: Humanity's First Visit to Our Star}},
	volume = {204},
	year = {2016}}

@article{Mueller2020,
	adsurl = {https://ui.adsabs.harvard.edu/abs/2020A&A...642A...1M},
	archiveprefix = {arXiv},
	author = {{M{\"u}ller}, D. and {St. Cyr}, O.~C. and {Zouganelis}, I. and {Gilbert}, H.~R. and {Marsden}, R. and {Nieves-Chinchilla}, T. and {Antonucci}, E. and {Auch{\`e}re}, F. and {Berghmans}, D. and {Horbury}, T.~S. and {Howard}, R.~A. and {Krucker}, S. and {Maksimovic}, M. and {Owen}, C.~J. and {Rochus}, P. and {Rodriguez-Pacheco}, J. and {Romoli}, M. and {Solanki}, S.~K. and {Bruno}, R. and {Carlsson}, M. and {Fludra}, A. and {Harra}, L. and {Hassler}, D.~M. and {Livi}, S. and {Louarn}, P. and {Peter}, H. and {Sch{\"u}hle}, U. and {Teriaca}, L. and {del Toro Iniesta}, J.~C. and {Wimmer-Schweingruber}, R.~F. and {Marsch}, E. and {Velli}, M. and {De Groof}, A. and {Walsh}, A. and {Williams}, D.},
	doi = {10.1051/0004-6361/202038467},
	eid = {A1},
	eprint = {2009.00861},
	journal = {\aap},
	month = oct,
	pages = {A1},
	primaryclass = {astro-ph.SR},
	title = {{The Solar Orbiter mission. Science overview}},
	volume = {642},
	year = 2020}

@article{Riley2012,
	adsurl = {http://adsabs.harvard.edu/abs/2012SoPh..277..355R},
	author = {{Riley}, P. and {Luhmann}, J.~G.},
	doi = {10.1007/s11207-011-9909-0},
	journal = {\solphys},
	month = apr,
	pages = {355-373},
	title = {{Interplanetary Signatures of Unipolar Streamers and the Origin of the Slow Solar Wind}},
	volume = 277,
	year = 2012}

@article{WangYM2000,
	adsurl = {http://adsabs.harvard.edu/abs/2000JGR...10525133W},
	author = {{Wang}, Y.-M. and {Sheeley}, N.~R. and {Socker}, D.~G. and {Howard}, R.~A. and {Rich}, N.~B.},
	doi = {10.1029/2000JA000149},
	journal = {\jgr},
	month = nov,
	pages = {25133-25142},
	title = {{The dynamical nature of coronal streamers}},
	volume = 105,
	year = 2000}

@ARTICLE{higginson_lynch_2018,
       author = {{Higginson}, A.~K. and {Lynch}, B.~J.},
        title = "{Structured Slow Solar Wind Variability: Streamer-blob Flux Ropes and Torsional Alfv{\'e}n Waves}",
      journal = {\apj},
         year = 2018,
        month = may,
       volume = {859},
       number = {1},
          eid = {6},
        pages = {6},
          doi = {10.3847/1538-4357/aabc08},
archivePrefix = {arXiv},
       eprint = {1710.00106},
 primaryClass = {astro-ph.SR},
       adsurl = {https://ui.adsabs.harvard.edu/abs/2018ApJ...859....6H}
}

@ARTICLE{reville_2020b,
       author = {{R{\'e}ville}, Victor and {Velli}, Marco and {Panasenco}, Olga and {Tenerani}, Anna and {Shi}, Chen and {Badman}, Samuel T. and {Bale}, Stuart D. and {Kasper}, J.~C. and {Stevens}, Michael L. and {Korreck}, Kelly E. and {Bonnell}, J.~W. and {Case}, Anthony W. and {de Wit}, Thierry Dudok and {Goetz}, Keith and {Harvey}, Peter R. and {Larson}, Davin E. and {Livi}, Roberto and {Malaspina}, David M. and {MacDowall}, Robert J. and {Pulupa}, Marc and {Whittlesey}, Phyllis L.},
        title = "{The Role of Alfv{\'e}n Wave Dynamics on the Large-scale Properties of the Solar Wind: Comparing an MHD Simulation with Parker Solar Probe E1 Data}",
      journal = {\apjs},
         year = 2020,
        month = feb,
       volume = {246},
       number = {2},
          eid = {24},
        pages = {24},
          doi = {10.3847/1538-4365/ab4fef},
archivePrefix = {arXiv},
       eprint = {1912.03777},
 primaryClass = {astro-ph.SR},
       adsurl = {https://ui.adsabs.harvard.edu/abs/2020ApJS..246...24R}
}

@ARTICLE{wang_evidence_2014,
       author = {{Wang}, Y. -M. and {Young}, P.~R. and {Muglach}, K.},
        title = "{Evidence for Two Separate Heliospheric Current Sheets of Cylindrical Shape During Mid-2012}",
      journal = {\apj},
         year = 2014,
        month = jan,
       volume = {780},
       number = {1},
          eid = {103},
        pages = {103},
          doi = {10.1088/0004-637X/780/1/103},
       adsurl = {https://ui.adsabs.harvard.edu/abs/2014ApJ...780..103W}
}

@article{reville_flux_2022,
	author = {{R{\'e}ville}, V. and {Fargette}, N. and {Rouillard}, A.~P. and {Lavraud}, B. and {Velli}, M. and {Strugarek}, A. and {Parenti}, S. and {Brun}, A.~S. and {Shi}, C. and {Kouloumvakos}, A. and {Poirier}, N. and {Pinto}, R.~F. and {Louarn}, P. and {Fedorov}, A. and {Owen}, C.~J. and {G{\'e}not}, V. and {Horbury}, T.~S. and {Laker}, R. and {O'Brien}, H. and {Angelini}, V. and {Fauchon-Jones}, E. and {Kasper}, J.~C.},
	doi = {10.1051/0004-6361/202142381},
	eid = {A110},
	eprint = {2112.07445},
	journal = {\aap},
	month = mar,
	pages = {A110},
	primaryclass = {astro-ph.SR},
	title = {{Flux rope and dynamics of the heliospheric current sheet. Study of the Parker Solar Probe and Solar Orbiter conjunction of June 2020}},
	volume = {659},
	year = 2022}

@ARTICLE{smith_heliospheric_2001,
	title = {The heliospheric current sheet},
	volume = {106},
	issn = {01480227},
	url = {http://doi.wiley.com/10.1029/2000JA000120},
	doi = {10.1029/2000JA000120},
	language = {en},
	number = {A8},
	urldate = {2023-09-17},
	journal = {Journal of Geophysical Research: Space Physics},
	author = {Smith, Edward J.},
	month = aug,
	year = {2001},
	pages = {15819--15831},
}

@ARTICLE{sokoloff_symmetries_2020,
	title = {Symmetries of {Magnetic} {Fields} {Driven} by {Spherical} {Dynamos} of {Exoplanets} and {Their} {Host} {Stars}},
	volume = {12},
	copyright = {http://creativecommons.org/licenses/by/3.0/},
	issn = {2073-8994},
	url = {https://www.mdpi.com/2073-8994/12/12/2085},
	doi = {10.3390/sym12122085},
	language = {en},
	number = {12},
	urldate = {2025-03-04},
	journal = {Symmetry},
	author = {Sokoloff, Dmitry and Malova, Helmi and Yushkov, Egor},
	month = dec,
	year = {2020},
	note = {Number: 12
Publisher: Multidisciplinary Digital Publishing Institute},
	pages = {2085},
}

@ARTICLE{sanches-diaz2019,
       author = {{Sanchez-Diaz}, E. and {Rouillard}, A.~P. and {Lavraud}, B. and {Kilpua}, E. and {Davies}, J.~A.},
        title = "{In Situ Measurements of the Variable Slow Solar Wind near Sector Boundaries}",
      journal = {\apj},
         year = 2019,
        month = sep,
       volume = {882},
       number = {1},
          eid = {51},
        pages = {51},
          doi = {10.3847/1538-4357/ab341c},
archivePrefix = {arXiv},
       eprint = {1911.09683},
 primaryClass = {astro-ph.SR},
       adsurl = {https://ui.adsabs.harvard.edu/abs/2019ApJ...882...51S}
}

@ARTICLE{mignone_2007,
       author = {{Mignone}, A. and {Bodo}, G. and {Massaglia}, S. and {Matsakos}, T. and {Tesileanu}, O. and {Zanni}, C. and {Ferrari}, A.},
        title = "{PLUTO: A Numerical Code for Computational Astrophysics}",
      journal = {\apjs},
         year = 2007,
        month = may,
       volume = {170},
       number = {1},
        pages = {228-242},
          doi = {10.1086/513316},
archivePrefix = {arXiv},
       eprint = {astro-ph/0701854},
 primaryClass = {astro-ph},
       adsurl = {https://ui.adsabs.harvard.edu/abs/2007ApJS..170..228M}
}

@ARTICLE{Rachmeler2014,
       author = {{Rachmeler}, L.~A. and {Platten}, S.~J. and {Bethge}, C. and {Seaton}, D.~B. and {Yeates}, A.~R.},
        title = "{Observations of a Hybrid Double-streamer/Pseudostreamer in the Solar Corona}",
      journal = {\apjl},
         year = 2014,
        month = may,
       volume = {787},
       number = {1},
          eid = {L3},
        pages = {L3},
          doi = {10.1088/2041-8205/787/1/L3},
archivePrefix = {arXiv},
       eprint = {1312.3153},
 primaryClass = {astro-ph.SR},
       adsurl = {https://ui.adsabs.harvard.edu/abs/2014ApJ...787L...3R}
}

@article{Panasenco2013,
	adsurl = {http://adsabs.harvard.edu/abs/2013AIPC.1539...50P},
	archiveprefix = {arXiv},
	author = {{Panasenco}, O. and {Velli}, M.},
	doi = {10.1063/1.4810987},
	eprint = {1211.6171},
	journal = {Solar Wind 13},
	month = jun,
	pages = {50-53},
	primaryclass = {astro-ph.SR},
	title = {{Coronal pseudostreamers: Source of fast or slow solar wind?}},
	volume = 1539,
	year = 2013}

@ARTICLE{wang_ps_2007,
       author = {{Wang}, Y.-M. and {Sheeley}, Jr., N.~R. and {Rich}, N.~B.},
        title = "{Coronal Pseudostreamers}",
      journal = {\apj},
         year = 2007,
        month = apr,
       volume = {658},
       number = {2},
        pages = {1340-1348},
          doi = {10.1086/511416},
       adsurl = {https://ui.adsabs.harvard.edu/abs/2007ApJ...658.1340W}
}

@ARTICLE{Athay1986,
       author = {{Athay}, R.~G.},
        title = "{Radiation Loss Rates in Lyman Alpha for Solar Conditions}",
      journal = {\apj},
         year = 1986,
        month = sep,
       volume = {308},
        pages = {975},
          doi = {10.1086/164565},
       adsurl = {https://ui.adsabs.harvard.edu/abs/1986ApJ...308..975A}
}

@ARTICLE{Alazraki1971,
       author = {{Alazraki}, G. and {Couturier}, P.},
        title = "{Solar Wind Acceleration Caused by the Gradient of Alfven Wave Pressure}",
      journal = {\aap},
         year = 1971,
        month = aug,
       volume = {13},
        pages = {380},
       adsurl = {https://ui.adsabs.harvard.edu/abs/1971A&A....13..380A}
}

@ARTICLE{Belcher1971,
       author = {{Belcher}, J.~W.},
        title = "{ALFV{\'E}NIC Wave Pressures and the Solar Wind}",
      journal = {\apj},
         year = 1971,
        month = sep,
       volume = {168},
        pages = {509},
          doi = {10.1086/151105},
       adsurl = {https://ui.adsabs.harvard.edu/abs/1971ApJ...168..509B}
}

@ARTICLE{Verdini2007,
       author = {{Verdini}, Andrea and {Velli}, Marco},
        title = "{Alfv{\'e}n Waves and Turbulence in the Solar Atmosphere and Solar Wind}",
      journal = {\apj},
         year = 2007,
        month = jun,
       volume = {662},
       number = {1},
        pages = {669-676},
          doi = {10.1086/510710},
archivePrefix = {arXiv},
       eprint = {astro-ph/0702205},
 primaryClass = {astro-ph},
       adsurl = {https://ui.adsabs.harvard.edu/abs/2007ApJ...662..669V}
}

@ARTICLE{Dedner2002,
       author = {{Dedner}, A. and {Kemm}, F. and {Kr{\"o}ner}, D. and {Munz}, C.-D. and {Schnitzer}, T. and {Wesenberg}, M.},
        title = "{Hyperbolic Divergence Cleaning for the MHD Equations}",
      journal = {Journal of Computational Physics},
         year = 2002,
        month = jan,
       volume = {175},
       number = {2},
        pages = {645-673},
          doi = {10.1006/jcph.2001.6961},
       adsurl = {https://ui.adsabs.harvard.edu/abs/2002JCoPh.175..645D}
}

@ARTICLE{vanderHolst2010,
       author = {{van der Holst}, B. and {Manchester}, IV, W.~B. and {Frazin}, R.~A. and {V{\'a}squez}, A.~M. and {T{\'o}th}, G. and {Gombosi}, T.~I.},
        title = "{A Data-driven, Two-temperature Solar Wind Model with Alfv{\'e}n Waves}",
      journal = {\apj},
         year = 2010,
        month = dec,
       volume = {725},
       number = {1},
        pages = {1373-1383},
          doi = {10.1088/0004-637X/725/1/1373},
       adsurl = {https://ui.adsabs.harvard.edu/abs/2010ApJ...725.1373V}
}

@ARTICLE{Hollweg1986,
       author = {{Hollweg}, J.~V.},
        title = "{Transition region, corona, and solar wind in coronal holes}",
      journal = {\jgr},
         year = 1986,
        month = apr,
       volume = {91},
       number = {A4},
        pages = {4111-4125},
          doi = {10.1029/JA091iA04p04111},
       adsurl = {https://ui.adsabs.harvard.edu/abs/1986JGR....91.4111H}
}

@ARTICLE{Reville2018,
       author = {{R{\'e}ville}, Victor and {Tenerani}, Anna and {Velli}, Marco},
        title = "{Parametric Decay and the Origin of the Low-frequency Alfv{\'e}nic Spectrum of the Solar Wind}",
      journal = {\apj},
         year = 2018,
        month = oct,
       volume = {866},
       number = {1},
          eid = {38},
        pages = {38},
          doi = {10.3847/1538-4357/aadb8f},
archivePrefix = {arXiv},
       eprint = {1806.05762},
 primaryClass = {astro-ph.SR},
       adsurl = {https://ui.adsabs.harvard.edu/abs/2018ApJ...866...38R}
}

@ARTICLE{Miyoshi2005,
       author = {{Miyoshi}, Takahiro and {Kusano}, Kanya},
        title = "{A multi-state HLL approximate Riemann solver for ideal magnetohydrodynamics}",
      journal = {Journal of Computational Physics},
         year = 2005,
        month = sep,
       volume = {208},
       number = {1},
        pages = {315-344},
          doi = {10.1016/j.jcp.2005.02.017},
       adsurl = {https://ui.adsabs.harvard.edu/abs/2005JCoPh.208..315M}
}

@ARTICLE{Reville2020ApJL,
       author = {{R{\'e}ville}, Victor and {Velli}, Marco and {Rouillard}, Alexis P. and {Lavraud}, Benoit and {Tenerani}, Anna and {Shi}, Chen and {Strugarek}, Antoine},
        title = "{Tearing Instability and Periodic Density Perturbations in the Slow Solar Wind}",
      journal = {\apjl},
         year = 2020,
        month = may,
       volume = {895},
       number = {1},
          eid = {L20},
        pages = {L20},
          doi = {10.3847/2041-8213/ab911d},
archivePrefix = {arXiv},
       eprint = {2005.02679},
 primaryClass = {astro-ph.SR},
       adsurl = {https://ui.adsabs.harvard.edu/abs/2020ApJ...895L..20R}
}

@article{Lynch2020,
	Adsurl = {https://ui.adsabs.harvard.edu/abs/2020ApJ...905..139L},
	Archiveprefix = {arXiv},
	Author = {{Lynch}, Benjamin J.},
	Doi = {10.3847/1538-4357/abc5b3},
	Eid = {139},
	Eprint = {2010.13959},
	Journal = {\apj},
	Month = dec,
	Number = {2},
	Pages = {139},
	Primaryclass = {astro-ph.SR},
	Title = {{A Model for Coronal Inflows and In/Out Pairs}},
	Volume = {905},
	Year = 2020}

@article{Viall2015,
	Adsurl = {http://adsabs.harvard.edu/abs/2015ApJ...807..176V},
	Author = {{Viall}, N.~M. and {Vourlidas}, A.},
	Doi = {10.1088/0004-637X/807/2/176},
	Eid = {176},
	Journal = {\apj},
	Month = jul,
	Pages = {176},
	Title = {{Periodic Density Structures and the Origin of the Slow Solar Wind}},
	Volume = 807,
	Year = 2015}

@article{Rouillard2010a,
	Adsurl = {http://adsabs.harvard.edu/abs/2010JGRA..115.4103R},
	Author = {{Rouillard}, A.~P. and {Davies}, J.~A. and {Lavraud}, B. and {Forsyth}, R.~J. and {Savani}, N.~P. and {Bewsher}, D. and {Brown}, D.~S. and {Sheeley}, N.~R. and {Davis}, C.~J. and {Harrison}, R.~A. and {Howard}, R.~A. and {Vourlidas}, A. and {Lockwood}, M. and {Crothers}, S.~R. and {Eyles}, C.~J.},
	Doi = {10.1029/2009JA014471},
	Eid = {A04103},
	Journal = {\jgr},
	Month = apr,
	Pages = {4103},
	Title = {{Intermittent release of transients in the slow solar wind: 1. Remote sensing observations}},
	Volume = 115,
	Year = 2010}

@article{Rouillard2010b,
	Adsurl = {http://adsabs.harvard.edu/abs/2010JGRA..115.4104R},
	Author = {{Rouillard}, A.~P. and {Lavraud}, B. and {Davies}, J.~A. and {Savani}, N.~P. and {Burlaga}, L.~F. and {Forsyth}, R.~J. and {Sauvaud}, J.-A. and {Opitz}, A. and {Lockwood}, M. and {Luhmann}, J.~G. and {Simunac}, K.~D.~C. and {Galvin}, A.~B. and {Davis}, C.~J. and {Harrison}, R.~A.},
	Doi = {10.1029/2009JA014472},
	Eid = {A04104},
	Journal = {\jgr},
	Month = apr,
	Pages = {4104},
	Title = {{Intermittent release of transients in the slow solar wind: 2. In situ evidence}},
	Volume = 115,
	Year = 2010}

@article{Sheeley1997,
	Adsurl = {http://adsabs.harvard.edu/abs/1997ApJ...484..472S},
	Author = {{Sheeley}, N.~R. and {Wang}, Y.-M. and {Hawley}, S.~H. and {Brueckner}, G.~E. and {Dere}, K.~P. and {Howard}, R.~A. and {Koomen}, M.~J. and {Korendyke}, C.~M. and {Michels}, D.~J. and {Paswaters}, S.~E. and {Socker}, D.~G. and {St.~Cyr}, O.~C. and {Wang}, D. and {Lamy}, P.~L. and {Llebaria}, A. and {Schwenn}, R. and {Simnett}, G.~M. and {Plunkett}, S. and {Biesecker}, D.~A.},
	Doi = {10.1086/304338},
	Journal = {\apj},
	Month = jul,
	Pages = {472-478},
	Title = {{Measurements of Flow Speeds in the Corona Between 2 and 30 R$_{\odot}$}},
	Volume = 484,
	Year = 1997}

@article{Sheeley2007,
	Adsurl = {http://adsabs.harvard.edu/abs/2007ApJ...655.1142S},
	Author = {{Sheeley}, Jr., N.~R. and {Wang}, Y.-M.},
	Doi = {10.1086/510323},
	Journal = {\apj},
	Month = feb,
	Pages = {1142-1156},
	Title = {{In/Out Pairs and the Detachment of Coronal Streamers}},
	Volume = 655,
	Year = 2007}

@article{Gannouni2023,
	adsurl = {https://ui.adsabs.harvard.edu/abs/2023ApJ...958..110G},
	archiveprefix = {arXiv},
	author = {{Gannouni}, Bahaeddine and {R{\'e}ville}, Victor and {Rouillard}, Alexis P.},
	doi = {10.3847/1538-4357/acfef3},
	eid = {110},
	eprint = {2307.02210},
	journal = {\apj},
	month = dec,
	number = {2},
	pages = {110},
	primaryclass = {astro-ph.SR},
	title = {{Modeling the Formation and Evolution of Solar Wind Microstreams: From Coronal Plumes to Propagating Alfv{\'e}nic Velocity Spikes}},
	volume = {958},
	year = 2023}

@article{Titov2011,
	adsurl = {http://adsabs.harvard.edu/abs/2011ApJ...731..111T},
	archiveprefix = {arXiv},
	author = {{Titov}, V.~S. and {Miki{\'c}}, Z. and {Linker}, J.~A. and {Lionello}, R. and {Antiochos}, S.~K.},
	doi = {10.1088/0004-637X/731/2/111},
	eid = {111},
	eprint = {1011.0009},
	journal = {\apj},
	month = apr,
	pages = {111},
	primaryclass = {astro-ph.SR},
	title = {{Magnetic Topology of Coronal Hole Linkages}},
	volume = 731,
	year = 2011}

@article{Titov2012,
	adsurl = {http://adsabs.harvard.edu/abs/2012ApJ...759...70T},
	archiveprefix = {arXiv},
	author = {{Titov}, V.~S. and {Mikic}, Z. and {T{\"o}r{\"o}k}, T. and {Linker}, J.~A. and {Panasenco}, O.},
	doi = {10.1088/0004-637X/759/1/70},
	eid = {70},
	eprint = {1209.5797},
	journal = {\apj},
	month = nov,
	pages = {70},
	primaryclass = {astro-ph.SR},
	title = {{2010 August 1-2 Sympathetic Eruptions. I. Magnetic Topology of the Source-surface Background Field}},
	volume = 759,
	year = 2012}

@article{Pellegrin-Frachon2023,
	adsurl = {https://ui.adsabs.harvard.edu/abs/2023A&A...675A..55P},
	author = {{Pellegrin-Frachon}, T. and {Masson}, S. and {Pariat}, {\'E}. and {Wyper}, P.~F. and {DeVore}, C.~R.},
	doi = {10.1051/0004-6361/202245611},
	eid = {A55},
	journal = {\aap},
	month = jul,
	pages = {A55},
	title = {{Interchange reconnection dynamics in a solar coronal pseudo-streamer}},
	volume = {675},
	year = 2023}

@article{Wyper2022,
	adsurl = {https://ui.adsabs.harvard.edu/abs/2022ApJ...941L..29W},
	author = {{Wyper}, Peter F. and {DeVore}, C.~R. and {Antiochos}, S.~K. and {Pontin}, D.~I. and {Higginson}, Aleida K. and {Scott}, Roger and {Masson}, Sophie and {Pelegrin-Frachon}, Theo},
	doi = {10.3847/2041-8213/aca8ae},
	eid = {L29},
	journal = {\apjl},
	month = dec,
	number = {2},
	pages = {L29},
	title = {{The Imprint of Intermittent Interchange Reconnection on the Solar Wind}},
	volume = {941},
	year = 2022}

@article{Romano2025,
	adsurl = {https://ui.adsabs.harvard.edu/abs/2025ApJ...982..142R},
	archiveprefix = {arXiv},
	author = {{Romano}, P. and {Wyper}, P. and {Andretta}, V. and {Antiochos}, S. and {Russano}, G. and {Spadaro}, D. and {Abbo}, L. and {Contarino}, L. and {Elmhamdi}, A. and {Ferrente}, F. and {Lionello}, R. and {Lynch}, B.~J. and {MacNeice}, P. and {Romoli}, M. and {Ventura}, R. and {Viall}, N. and {Bemporad}, A. and {Burtovoi}, A. and {Da Deppo}, V. and {De Leo}, Y. and {Fineschi}, S. and {Frassati}, F. and {Giordano}, S. and {Guglielmino}, S.~L. and {Grimani}, C. and {Heinzel}, P. and {Jerse}, G. and {Landini}, F. and {Naletto}, G. and {Pancrazzi}, M. and {Sasso}, C. and {Stangalini}, M. and {Susino}, R. and {Telloni}, D. and {Teriaca}, L. and {Uslenghi}, M.},
	doi = {10.3847/1538-4357/adb1da},
	eid = {142},
	eprint = {2502.08015},
	journal = {\apj},
	month = apr,
	number = {2},
	pages = {142},
	primaryclass = {astro-ph.SR},
	title = {{Metis Observations of Alfv{\'e}nic Outflows Driven by Interchange Reconnection in a Pseudostreamer}},
	volume = {982},
	year = 2025}

@article{Masson2014,
	adsurl = {http://adsabs.harvard.edu/abs/2014ApJ...787..145M},
	archiveprefix = {arXiv},
	author = {{Masson}, S. and {McCauley}, P. and {Golub}, L. and {Reeves}, K.~K. and {DeLuca}, E.~E.},
	doi = {10.1088/0004-637X/787/2/145},
	eid = {145},
	eprint = {1301.0740},
	journal = {\apj},
	month = jun,
	pages = {145},
	primaryclass = {astro-ph.SR},
	title = {{Dynamics of the Transition Corona}},
	volume = 787,
	year = 2014}

@article{Lynch2013,
	adsurl = {http://adsabs.harvard.edu/abs/2013ApJ...764...87L},
	archiveprefix = {arXiv},
	author = {{Lynch}, B.~J. and {Edmondson}, J.~K.},
	doi = {10.1088/0004-637X/764/1/87},
	eid = {87},
	eprint = {1212.6677},
	journal = {\apj},
	month = feb,
	pages = {87},
	primaryclass = {astro-ph.SR},
	title = {{Sympathetic Magnetic Breakout Coronal Mass Ejections from Pseudostreamers}},
	volume = 764,
	year = 2013}

@article{Lynch2014,
	adsurl = {http://adsabs.harvard.edu/abs/2014SoPh..289.3043L},
	archiveprefix = {arXiv},
	author = {{Lynch}, B.~J. and {Edmondson}, J.~K. and {Li}, Y.},
	doi = {10.1007/s11207-014-0506-x},
	eprint = {1401.7965},
	journal = {\solphys},
	month = aug,
	pages = {3043-3058},
	primaryclass = {astro-ph.SR},
	title = {{Interchange Reconnection Alfv{\'e}n Wave Generation}},
	volume = 289,
	year = 2014}

@article{Lynch2023,
	adsurl = {https://ui.adsabs.harvard.edu/abs/2023ApJ...949...14L},
	archiveprefix = {arXiv},
	author = {{Lynch}, B.~J. and {Viall}, N.~M. and {Higginson}, A.~K. and {Zhao}, L. and {Lepri}, S.~T. and {Sun}, X.},
	doi = {10.3847/1538-4357/acc38c},
	eid = {14},
	eprint = {2303.06465},
	journal = {\apj},
	month = may,
	number = {1},
	pages = {14},
	primaryclass = {astro-ph.SR},
	title = {{The S-Web Origin of Composition Enhancement in the Slow-to-moderate Speed Solar Wind}},
	volume = {949},
	year = 2023}

@article{WangYM2012,
	adsurl = {http://adsabs.harvard.edu/abs/2012ApJ...749..182W},
	author = {{Wang}, Y.-M. and {Grappin}, R. and {Robbrecht}, E. and {Sheeley}, Jr., N.~R.},
	doi = {10.1088/0004-637X/749/2/182},
	eid = {182},
	journal = {\apj},
	month = apr,
	pages = {182},
	title = {{On the Nature of the Solar Wind from Coronal Pseudostreamers}},
	volume = 749,
	year = 2012}

@article{WangYM2008,
	adsurl = {http://adsabs.harvard.edu/abs/2008SoPh..249...17W},
	author = {{Wang}, Y.-M. and {Muglach}, K.},
	doi = {10.1007/s11207-008-9171-2},
	journal = {\solphys},
	month = may,
	pages = {17-35},
	title = {{Observations of Low-Latitude Coronal Plumes}},
	volume = 249,
	year = 2008}

@article{WangYM2019,
	adsurl = {https://ui.adsabs.harvard.edu/abs/2019ApJ...872..139W},
	author = {{Wang}, Y. -M. and {Panasenco}, O.},
	doi = {10.3847/1538-4357/aaff5e},
	eid = {139},
	journal = {\apj},
	month = feb,
	number = {2},
	pages = {139},
	title = {{Observations of Solar Wind from Earth-directed Coronal Pseudostreamers}},
	volume = {872},
	year = 2019}

@article{Demoulin1996,
	adsurl = {https://ui.adsabs.harvard.edu/abs/1996JGR...101.7631D},
	author = {{D{\'e}moulin}, P. and {Priest}, E.~R. and {Lonie}, D.~P.},
	doi = {10.1029/95JA03558},
	journal = {\jgr},
	month = apr,
	number = {A4},
	pages = {7631-7646},
	title = {{Three-dimensional magnetic reconnection without null points 2. Application to twisted flux tubes}},
	volume = {101},
	year = 1996}

@ARTICLE{Antiochos1990,
       author = {{Antiochos}, Spiro K.},
        title = "{Heating of the corona by magnetic singularities}",
      journal = {\memsai},
         year = 1990,
        month = jan,
       volume = {61},
       number = {2},
        pages = {369-382},
       adsurl = {https://ui.adsabs.harvard.edu/abs/1990MmSAI..61..369A}
}

@article{Edmondson2017,
	adsurl = {https://ui.adsabs.harvard.edu/#abs/2017ApJ...849...28E},
	author = {{Edmondson}, J.~K. and {Lynch}, B.~J.},
	doi = {10.3847/1538-4357/aa83ba},
	eid = {28},
	journal = {\apj},
	month = Nov,
	pages = {28},
	primaryclass = {astro-ph.SR},
	title = {{Formation and Reconnection of Three-dimensional Current Sheets with a Guide Field in the Solar Corona}},
	volume = {849},
	year = 2017}

@article{Charbonneau2020,
	adsurl = {https://ui.adsabs.harvard.edu/abs/2020LRSP...17....4C},
	author = {{Charbonneau}, Paul},
	doi = {10.1007/s41116-020-00025-6},
	eid = {4},
	journal = {Living Reviews in Solar Physics},
	month = dec,
	number = {1},
	pages = {4},
	title = {{Dynamo models of the solar cycle}},
	volume = {17},
	year = 2020}

@ARTICLE{Hathaway2015,
       author = {{Hathaway}, David H.},
        title = "{The Solar Cycle}",
      journal = {Living Reviews in Solar Physics},
         year = 2015,
        month = dec,
       volume = {12},
       number = {1},
          eid = {4},
        pages = {4},
          doi = {10.1007/lrsp-2015-4},
archivePrefix = {arXiv},
       eprint = {1502.07020},
 primaryClass = {astro-ph.SR},
       adsurl = {https://ui.adsabs.harvard.edu/abs/2015LRSP...12....4H}
}

@ARTICLE{Badman2025,
       author = {{Badman}, Samuel T. and {Stevens}, Michael L. and {Bale}, Stuart D. and {Rivera}, Yeimy J. and {Klein}, Kristopher G. and {Niembro}, Tatiana and {Chhiber}, Rohit and {Rahmati}, Ali and {Whittlesey}, Phyllis L. and {Livi}, Roberto and {Larson}, Davin E. and {Owen}, Christopher J. and {Paulson}, Kristoff W. and {Horbury}, Timothy S. and {Morris}, Jean and {O'Brien}, Helen and {Dakeyo}, Jean-Baptiste and {Verniero}, Jaye L. and {Martinovic}, Mihailo and {Pulupa}, Marc and {Fraschetti}, Federico},
        title = "{Multispacecraft Measurements of the Evolving Geometry of the Solar Alfv{\'e}n Surface over Half a Solar Cycle}",
      journal = {\apjl},
         year = 2025,
        month = dec,
       volume = {995},
       number = {2},
          eid = {L37},
        pages = {L37},
          doi = {10.3847/2041-8213/ae0e5c},
archivePrefix = {arXiv},
       eprint = {2509.17149},
 primaryClass = {astro-ph.SR},
       adsurl = {https://ui.adsabs.harvard.edu/abs/2025ApJ...995L..37B}
}

@article{Vourlidas2016,
	adsurl = {https://ui.adsabs.harvard.edu/abs/2016SSRv..204...83V},
	author = {{Vourlidas}, Angelos and {Howard}, Russell A. and {Plunkett}, Simon P. and {Korendyke}, Clarence M. and {Thernisien}, Arnaud F.~R. and {Wang}, Dennis and {Rich}, Nathan and {Carter}, Michael T. and {Chua}, Damien H. and {Socker}, Dennis G. and {Linton}, Mark G. and {Morrill}, Jeff S. and {Lynch}, Sean and {Thurn}, Adam and {Van Duyne}, Peter and {Hagood}, Robert and {Clifford}, Greg and {Grey}, Phares J. and {Velli}, Marco and {Liewer}, Paulett C. and {Hall}, Jeffrey R. and {DeJong}, Eric M. and {Mikic}, Zoran and {Rochus}, Pierre and {Mazy}, Emanuel and {Bothmer}, Volker and {Rodmann}, Jens},
	doi = {10.1007/s11214-014-0114-y},
	journal = {\ssr},
	month = dec,
	number = {1-4},
	pages = {83-130},
	title = {{The Wide-Field Imager for Solar Probe Plus (WISPR)}},
	volume = {204},
	year = 2016}

@ARTICLE{Vourlidas2025,
       author = {{Vourlidas}, Angelos and {Paouris}, Evangelos and {Linton}, Mark G. and {Stenborg}, Guillermo and {Liewer}, Paulett C. and {Riley}, Pete and {Hess}, Phillip and {Howard}, Russell A. and {Raouafi}, Nour E.},
        title = "{High-resolution Imaging of the Magnetic Reconfiguration of the Corona from inside the Corona by WISPR on Parker Solar Probe}",
      journal = {\apjl},
         year = 2025,
        month = dec,
       volume = {995},
       number = {2},
          eid = {L38},
        pages = {L38},
          doi = {10.3847/2041-8213/ae0d7d},
       adsurl = {https://ui.adsabs.harvard.edu/abs/2025ApJ...995L..38V}
}

@ARTICLE{Cappello2024,
       author = {{Cappello}, G.~M. and {Temmer}, M. and {Vourlidas}, A. and {Braga}, C. and {Liewer}, P.~C. and {Qiu}, J. and {Stenborg}, G. and {Kouloumvakos}, A. and {Veronig}, A.~M. and {Bothmer}, V.},
        title = "{Internal magnetic field structures observed by PSP/WISPR in a filament-related coronal mass ejection}",
      journal = {\aap},
         year = 2024,
        month = aug,
       volume = {688},
          eid = {A162},
        pages = {A162},
          doi = {10.1051/0004-6361/202449613},
archivePrefix = {arXiv},
       eprint = {2402.14682},
 primaryClass = {astro-ph.SR},
       adsurl = {https://ui.adsabs.harvard.edu/abs/2024A&A...688A.162C}
}

@ARTICLE{Ascione2024,
       author = {{Ascione}, Madison L. and {Gutarra-Leon}, Angel J. and {Shaik}, Shaheda Begum and {Linton}, Mark G. and {Battams}, Karl and {Liewer}, Paulett C. and {Gallagher}, Brendan M.},
        title = "{A Detailed Analysis of a Magnetic Island Observed by WISPR on Parker Solar Probe}",
      journal = {\apj},
         year = 2024,
        month = sep,
       volume = {973},
       number = {1},
          eid = {12},
        pages = {12},
          doi = {10.3847/1538-4357/ad5e76},
archivePrefix = {arXiv},
       eprint = {2407.07216},
 primaryClass = {astro-ph.SR},
       adsurl = {https://ui.adsabs.harvard.edu/abs/2024ApJ...973...12A}
}

@ARTICLE{Liewer2024,
       author = {{Liewer}, Paulett C. and {Gallagher}, Brendan M. and {Stenborg}, Guillermo and {Linton}, Mark G. and {Qiu}, Jiong and {Vourlidas}, Angelos and {Ascione}, Madison L. and {Velli}, Marco},
        title = "{Evidence of Continuous Reconnection along a Helmet Streamer Current Sheet Observed by WISPR on Parker Solar Probe}",
      journal = {\apj},
         year = 2024,
        month = jul,
       volume = {970},
       number = {1},
          eid = {79},
        pages = {79},
          doi = {10.3847/1538-4357/ad509b},
       adsurl = {https://ui.adsabs.harvard.edu/abs/2024ApJ...970...79L}
}

@article{Liewer2023,
	adsurl = {https://ui.adsabs.harvard.edu/abs/2023ApJ...948...24L},
	author = {{Liewer}, Paulett C. and {Vourlidas}, Angelos and {Stenborg}, Guillermo and {Howard}, Russell A. and {Qiu}, Jiong and {Penteado}, Paulo and {Panasenco}, Olga and {Braga}, Carlos R.},
	doi = {10.3847/1538-4357/acc8c7},
	eid = {24},
	journal = {\apj},
	month = may,
	number = {1},
	pages = {24},
	title = {{Structure of the Plasma near the Heliospheric Current Sheet as Seen by WISPR/Parker Solar Probe from inside the Streamer Belt}},
	volume = {948},
	year = 2023}

@article{Zhukov2008,
	adsurl = {https://ui.adsabs.harvard.edu/abs/2008ApJ...680.1532Z},
	author = {{Zhukov}, A.~N. and {Saez}, F. and {Lamy}, P. and {Llebaria}, A. and {Stenborg}, G.},
	doi = {10.1086/587924},
	journal = {\apj},
	month = jun,
	number = {2},
	pages = {1532-1541},
	title = {{The Origin of Polar Streamers in the Solar Corona}},
	volume = {680},
	year = 2008}

@article{Lynch2016b,
	author = {{Lynch}, B.~J. and {Masson}, S. and {Li}, Y. and {Devore}, C.~R. and {Luhmann}, J.~G. and {Antiochos}, S.~K. and {Fisher}, G.~H.},
	doi = {10.1002/2016JA023432},
	eid = {10677},
	journal = {\jgr},
	month = nov,
	pages = {10677},
	title = {{A model for stealth coronal mass ejections}},
	volume = 121,
	year = 2016
}

@ARTICLE{Winterhalter94,
       author = {{Winterhalter}, D. and {Smith}, E.~J. and {Burton}, M.~E. and {Murphy}, N. and {McComas}, D.~J.},
        title = "{The heliospheric plasma sheet}",
      journal = {\jgr},
         year = 1994,
        month = apr,
       volume = {99},
       number = {A4},
        pages = {6667-6680},
          doi = {10.1029/93JA03481},
       adsurl = {https://ui.adsabs.harvard.edu/abs/1994JGR....99.6667W}
}

@ARTICLE{Mursula12,
       author = {{Mursula}, K. and {Virtanen}, I.~I.},
        title = "{The wide skirt of the bashful ballerina: Hemispheric asymmetry of the heliospheric magnetic field in the inner and outer heliosphere}",
      journal = {Journal of Geophysical Research (Space Physics)},
         year = 2012,
        month = aug,
       volume = {117},
       number = {A8},
          eid = {A08104},
        pages = {A08104},
          doi = {10.1029/2011JA017197},
       adsurl = {https://ui.adsabs.harvard.edu/abs/2012JGRA..117.8104M}
}

@article{McComasetal07,
    author = {McComas, D. J. and Velli, M. and Lewis, W. S. and Acton, L. W. and Balat-Pichelin, M. and Bothmer, V. and Dirling Jr., R. B. and Feldman, W. C. and Gloeckler, G. and Habbal, S. R. and Hassler, D. M. and Mann, I. and Matthaeus, W. H. and McNutt Jr., R. L. and Mewaldt, R. A. and Murphy, N. and Ofman, L. and Sittler Jr., E. C. and Smith, C. W. and Zurbuchen, T. H.},
    title = {Understanding coronal heating and solar wind acceleration: Case for in situ near-Sun measurements},
    journal = {Reviews of Geophysics},
    volume = {45},
    number = {1},
    pages = {},
    doi = {https://doi.org/10.1029/2006RG000195},
    url = {https://agupubs.onlinelibrary.wiley.com/doi/abs/10.1029/2006RG000195},
    eprint = {https://agupubs.onlinelibrary.wiley.com/doi/pdf/10.1029/2006RG000195},
    year = {2007}
}

@ARTICLE{Caplan2025,
       author = {{Caplan}, Ronald M. and {Stulajter}, Miko M. and {Linker}, Jon A. and {Downs}, Cooper and {Upton}, Lisa A. and {Jha}, Bibhuti Kumar and {Attie}, Raphael and {Arge}, Charles N. and {Henney}, Carl J.},
        title = "{Open-source Flux Transport (OFT). I. HipFT─High-performance Flux Transport}",
      journal = {\apjs},
         year = 2025,
        month = may,
       volume = {278},
       number = {1},
          eid = {24},
        pages = {24},
          doi = {10.3847/1538-4365/adc080},
archivePrefix = {arXiv},
       eprint = {2501.06377},
 primaryClass = {astro-ph.SR},
       adsurl = {https://ui.adsabs.harvard.edu/abs/2025ApJS..278...24C}
}

@article{Panasenco2019,
	Adsurl = {https://ui.adsabs.harvard.edu/abs/2019ApJ...873...25P},
	Author = {{Panasenco}, Olga and {Velli}, Marco and {Panasenco}, Aram},
	Doi = {10.3847/1538-4357/ab017c},
	Eid = {25},
	Journal = {\apj},
	Month = mar,
	Number = {1},
	Pages = {25},
	Title = {{Large-scale Magnetic Funnels in the Solar Corona}},
	Volume = {873},
	Year = 2019}

@article{Linker2003,
	Adsurl = {http://adsabs.harvard.edu/abs/2003PhPl...10.1971L},
	Author = {{Linker}, J.~A. and {Miki{\'c}}, Z. and {Lionello}, R. and {Riley}, P. and {Amari}, T. and {Odstrcil}, D.},
	Doi = {10.1063/1.1563668},
	Journal = {Physics of Plasmas},
	Month = may,
	Pages = {1971-1978},
	Title = {{Flux cancellation and coronal mass ejections}},
	Volume = 10,
	Year = 2003}

@article{Janvier2015,
	Adsurl = {http://adsabs.harvard.edu/abs/2015SoPh..290.3425J},
	Archiveprefix = {arXiv},
	Author = {{Janvier}, M. and {Aulanier}, G. and {D{\'e}moulin}, P.},
	Doi = {10.1007/s11207-015-0710-3},
	Eprint = {1505.05299},
	Journal = {\solphys},
	Month = dec,
	Pages = {3425-3456},
	Primaryclass = {astro-ph.SR},
	Title = {{From Coronal Observations to MHD Simulations, the Building Blocks for 3D Models of Solar Flares (Invited Review)}},
	Volume = 290,
	Year = 2015}

\end{document}